%% file: rel_charm.tex
\documentstyle[prd,aps,eqsecnum,preprint, tighten,epsf,epsfig]{revtex}
\tightenlines

\newcommand{\be}{\begin{equation}}
\newcommand{\ee}{\end{equation}}
\newcommand{\bea}{\begin{eqnarray}}
\newcommand{\eea}{\end{eqnarray}}
\newcommand{\bean}{\begin{eqnarray*}}
\newcommand{\eean}{\end{eqnarray*}}

\newcommand{\g}{\gamma}

\newcommand{\e}{\epsilon}

\begin{document}

\preprint{CU-TP-1063}

\title{Excited charmonium spectrum from anisotropic lattices}
\author{ X.~Liao and T.~Manke }
\address{Physics Department, Columbia University, New York, NY 10027, USA}

\maketitle
 
\begin{abstract}
We present our final results for the excited charmonium spectrum from a
quenched calculation using a fully relativistic anisotropic lattice QCD 
action. 
A detailed excited charmonium spectrum is obtained, including both the exotic hybrids 
(with $J^{PC} = 1^{-+}, 0^{+-}, 2^{+-}$) and orbitally excited mesons 
(with orbital angular momentum up to 3). Using three 
different lattice spacings (0.197, 0.131, and 0.092 fm), 
we perform a continuum extrapolation of the 
spectrum. We convert our results in lattice units to physical values 
using lattice scales set by the $^1P_1-1S$ splitting. 
The lowest lying exotic hybrid $1^{-+}$ lies at 4.428(41) GeV, slightly above the 
$D^{**}D$ (S+P wave) threshold of 4.287 GeV. Another two exotic hybrids $0^{+-}$ and $2^{+-}$ 
are determined to be 4.70(17) GeV and 4.895(88) GeV, respectively. Our finite volume analysis
confirms that our lattices are large enough to accommodate all the
excited states reported here. 
\end{abstract}

\section{Introduction}
\label{sec:intro}
The hadron spectrum is one of the most prominent non-perturbative consequences of QCD.
Over the past two decades, enormous progress has been achieved in developing the non-perturbative
techniques needed to predict these hadron masses from first principles. 
It is crucial that our theoretical techniques  reproduce the observed hadron spectrum 
to justify predictions for 
more complicated quantities such as the weak matrix elements.

On the experimental front, data for the excited charmonium spectrum above the $D\bar{D}$ threshold
are scarce, and no heavy hybrid signal has been observed yet, even though 
some potential candidates for light exotic hybrids  have been found and discussed extensively 
\cite{hybrid-review:page}. However, the situation will change dramatically in the next few years
with more data from the $B$-factories PEP-II (Babar) and KEKB (Belle). Even more important results
are promised by the focused charm physics 
programs proposed for the upgraded CLEO-c detector\cite{cleo-C:stock2002} (expected to start running in 
early 2003) and the new BESIII detector \cite{BES-charm:2001} (expected in 2005-2006) at BEPC
(Beijing Electron-Position Collider). Both CLEO-c and BESIII will collect large amounts of data for 
the charmonium spectrum above the $D\bar{D}$ threshold and provide badly needed, accurate results 
for the excited charmonium
spectrum, including radially and orbitally excited conventional $c\bar{c}$ mesons and gluon-rich 
states (glueballs and hybrids). Better theoretical predictions, especially those from lattice QCD 
calculations, will provide important guidance for these experimental efforts.  

In contrast to the light hadron spectrum, the heavy quark spectrum is
much better understood theoretically with various phenomenological models 
such as the static potential models 
and, more reliably, with the low energy effective theories such
as non-relativistic QCD (NRQCD) \cite{nrqcd:improve,heavy:Davies}. NRQCD has been
the dominant approach to the study of  heavy quark physics for many years and has been quite successful,
especially for the bottomonium systems. However, it is rather difficult to 
control the systematic errors of NRQCD, including the relativistic corrections and 
the radiative corrections, which  were shown to be quite large for charmonium 
\cite{nrqcd:trottier-rel,nrqcd:trottier-spin},
and still sizable for bottomonium \cite{nrqcd:manke-bottom-1997,nrqcd:Eicker-bottom-1998}.
Of even greater concern are the finite lattice spacing artifacts that could not be controlled by 
taking the continuum limit because of the restriction: $M_q a > 1$. 
Nonetheless, all these approximations to QCD provide important guidance and calibration for 
fully relativistic lattice QCD calculations and are quite useful 
for the understanding of systematic errors such as those arising from quenching and
finite volume effects.

Our ultimate goal is to determine the heavy quark spectrum (extrapolated to the continuum
limit) from fully relativistic 
lattice QCD. However, the large separation of energy scales ($m_q v^2 \ll m_q v \ll m_q$) in 
the heavy quark system makes it prohibitively expensive 
to study heavy quarks on a 
conventional isotropic lattice. Earlier relativistic
studies employed the ``Fermilab approach'' \cite{Fermilab,Fermilab:ukqcdmeson}, which uses a 
space-time asymmetric, ${\cal O}(a)$ improved action to control the ${\cal O}(M_q a)$ errors on an isotropic
lattice with equal lattice spacings in the space and time direction ($a_t = a_s$). In this study, 
we employ a newer approach called relativistic anisotropic lattice QCD, which is a generalization
of the ``Fermilab action'' to an anisotropic lattice with $a_t < a_s$. The anisotropic lattice gauge
action was developed in the 1980s \cite{aniso:karsch},  and has been
used quite successfully for glueball spectrum calculations
\cite{aniso:glueball:1997,aniso:glueball:1999} and thermodynamic studies \cite{aniso:thermo-Namekawa2001}. 
However, it was not until the late 1990s that an 
improved relativistic anisotropic fermion action was implemented for 
Wilson-type quarks \cite{aniso:klassen-improv1997,aniso:klassen-latproc}. Since then, the relativistic 
anisotropic lattice technique has been applied to calculate the heavy quark spectrum with unprecedented
accuracy,  including both charmonium \cite{aniso:klassen-latproc,aniso:pchen,aniso:cppacs-charm-lat00} 
and bottomonium states \cite{own-bottom-paper}.

 In addition to conventional mesons made of a quark-antiquark pair (within the naive
quark model), hybrid mesons containing valence gluons are also believed to be present in the nature 
\footnote{We use $q\bar{q}$ to represent a conventional meson (a pure quarkonium)
	  and $q\bar{q}g$ for a hybrid meson.
}. 
Phenomenological studies using the Born-Oppenheimer approximation \cite{hybrid:bornapprox}, the bag model 
\cite{bagmodel:barnes,bagmodel:charm}, the flux-tube model 
\cite{flux_tube}, and the QCD sum rules \cite{sumrule:1982,sumrule:heavy}, have provided 
the theoretical evidence and given semi-quantitative estimates for these hybrid states,
which are further confirmed by NRQCD \cite{NRQCDHybrids-ukqcd} and standard lattice QCD calculations \cite{milc_hybrid}. 
In the naive quark model, not all $J^{PC}$ quantum numbers are available, since
the parity (P) and charge conjugation (C) of a conventional $q\bar{q}$ meson are given by:
\be
	P=(-1)^{L+1}, C=(-1)^{L+S},
\label{qq_jpc}
\ee
where $L$ and $S$ are the orbital angular momentum and spin of the $q\bar{q}$ pair.
Hybrid states possessing so-called exotic quantum numbers (eg. $1^{-+}$, $0^{+-}$, and $2^{+-}$) 
have attracted considerable attention because such states do not mix with  the conventional
$q\bar{q}$ mesons and therefore should be more easily distinguished in experiments than can 
non-exotic hybrids.
More comprehensive reviews of hybrid excitations can be found in Ref. \cite{hybrid-review:page,exotic-review-Kuti1998}.

However, like all the other highly excited states, these hybrid mesons are difficult to 
measure accurately on a lattice because their
signals are much nosier than those of the low lying mesons.
Thus, it is crucial to have fine temporal resolution as well as highly optimized operators
to collect as much information as possible from meson correlation functions even at 
short time seperation.
The anisotropic lattice technique provides an efficient framework for such high 
precision calculations, because  it allows independent control of the temporal and 
spacial lattice spacings, thereby permitting simulations with a very fine grid
in the temporal direction.

In this work, we employ well optimized extended operators to study the excited charmonium
spectrum, including 
the orbitally excited mesons (P-, D-, and F-wave) and hybrid mesons 
with exotic quantum numbers. This complements our group's previous charmonium calculations 
on anisotropic lattices \cite{aniso:pchen}, which used only local operators and focused on
the conventional low lying S-wave and P-wave $c\bar{c}$ mesons. This paper is a follow-up
and complete analysis of our earlier work \cite{own-lat00}.

This paper is organized as follows. Sec. \ref{sec:actions} describes the anisotropic
lattice gauge action and fermion action. A discussion of the construction and optimization
of meson operators is given in section \ref{sec:meson_meas}. Sec. \ref{sec:simulation}
contains the details of our simulations, including simulation parameters and some measures taken
to improve computational efficiency. In section \ref{sec:results}, we present our spectrum
results and discuss the systematic errors and the theoretical and experimental implications.
Section \ref{sec:conclusion} contains concluding remarks. In the appendix,
we describe the notation used in this paper, our meson naming convention, and some relevant 
properties of the hyper-cubic symmetry group.

\section{Anisotropic lattice action}
\label{sec:actions}

We employ an anisotropic gluon action \cite{aniso:karsch}, which is accurate up to 
${\cal O}(a_s^2,a_t^2)$ discretization errors:

\bea
S = - \beta \left( \xi_0^{-1} \sum_{x, {\rm i > j}} P_{\rm i j}(x) +
    \xi_0  \sum_{x, {\rm i}} P_{\rm i 0} (x) \right) ~~.
\label{eq:aniso_glue}
\eea
This is the standard Wilson action written in terms of simple plaquettes,
$P_{\mu\nu}(x)$. Here $0$ labels the time component and the index $i=1,2,3$ runs over
the three spatial directions. The parameters $\beta$ and $\xi_0$ are the bare coupling and bare anisotropy respectively, which 
determine the spatial lattice spacing, $a_s$, and the renormalized 
anisotropy, $\xi$, of the quenched lattice. The continuum limit should be taken at 
fixed anisotropy, $\xi = a_s/a_t$.
For the heavy quark propagation in the gluon background 
we used the ``anisotropic clover'' formulation as first described in 
\cite{aniso:klassen-latproc}. The discretized form of the continuum Dirac operator, 
$Q=m_q+D\hskip -0.21cm \slash $, reads
\bea
Q  & = &  m_0 + \nu_s~W_i \gamma_i + \nu_t~W_0 \gamma_0 - \nonumber 
\frac{a_s}{2}\left[ c_t~\sigma_{0k}F_{0k} + c_s~\sigma_{kl}F_{kl} \right]~~, 
\nonumber \\
W_\mu & = & \nabla_\mu - (a_\mu/2) \gamma_\mu \Delta_\mu ~~.
\label{eq:aniso_quark}
\eea
Here $\nabla_\mu$ is the symmetric lattice derivative:
\bea
\bar q(x)~a_\mu\nabla_\mu ~q(x) &=& \bar q(x)~U_\mu(x) ~q(x+\mu) - \bar q(x)~U_{-\mu}(x)~ q(x-\mu).
\label{eq:lat_deriv_op} 
\eea
and $\Delta_\mu$ is defined by $a_\mu^2 \Delta_\mu q(x) \equiv U_\mu(x)~q(x+\mu) -2~q(x) + U_{-\mu}(x)~q(x-\mu)$.
For the field tensor $F_{\mu \nu}$ we choose the traceless cloverleaf
definition which sums the 4 plaquettes centered at point $x$ in the $(\mu, \nu)$ plane:
\be
a_\mu a_\nu~F_{\mu \nu}(x) \equiv \frac{i}{2} \left[ P_{\mu \nu}(x) + P_{\nu \bar \mu}(x) +  P_{\bar \mu  \bar \nu}(x) + P_{\bar \nu \mu}(x) - h.c. \right]~~,
\ee
where $\bar{\mu}$ indicates the negative direction $-\hat{\mu}$, so $P_{\bar \mu  \bar \nu}(x)$ represents $U^{\dag}(x-\hat{\mu},\mu)U^{\dag}(x-\hat{\mu}-\hat{\nu},\nu)U(x-\hat{\mu}-\hat{\nu},\mu)U(x-\hat{\nu},\nu)$.
This is indeed the most general anisotropic quark action 
including all operators to dimension 5 up to $O(3)$ symmetric field 
redefinition. 
We have chosen Wilson's combination, $W_\mu$, of first and second derivative 
terms so as to both remove all doublers and ensure the full projection property.
The five parameters in Eq. (\ref{eq:aniso_quark}) are all related
to the quark mass, $m_q$, and the gauge coupling as they appear in the
continuum action. By tuning them appropriately we can remove all ${\cal O}(a)$ errors and re-establish space-time axis-interchange symmetry for long-distance physics.
A more detailed discussion of these parameters can be found in Refs
\cite{aniso:klassen-latproc,aniso:pchen,own-bottom-paper,aniso:aoki}.

Classical values for these parameters have been given in Ref.~\cite{aniso:pchen}: \\
\bea
m_0   &=& m_q(1 + \frac{1}{2}~a_sm_q) \\
\nu_t &=& \nu_s \frac{1 + \frac{1}{2}~a_s m_q}{1 + \frac{1}{2}~a_t m_q} \\
c_s   &=& \nu_s \label{eq:c_s} \\
c_t   &=& \frac{1}{2}\left(\nu_s + \nu_t \frac{a_t}{a_s}\right)~~\label{eq:c_t}.
\eea
Simple field rescaling enables us to set one of the above coefficients 
at will. For convenience we fix $\nu_s=1$ and adjust $\nu_t$ 
non-perturbatively requiring that the mesons obey a relativistic dispersion relation 
(~$c({\bf 0})=1$~):
\bea
E^2({\bf p}) &=& E^2({\bf 0}) + c^2({\bf p})~{\bf p}^2 + {\cal O} ({\bf p}^4) \ldots ~~.
\eea
We also choose $m_0$ non-perturbatively, such that the spin average of the lowest S-wave
mesons ($1S$) matches its experimental value: \mbox{$\frac{3}{4}M(^{3}S_1) + \frac{1}{4}M(^{1}S_0)$ =} \mbox{3.067 GeV} for charmonium.
For the clover coefficients $(c_s,c_t)$ we take their classical estimates
from Eqs. (\ref{eq:c_s}) and (\ref{eq:c_t}) and augment them 
by tadpole improvement. 
\bea
c_s &\to& c_s/(u_s^3) \\
c_t &\to& c_t/(u_t u_s^2)
\eea
The tadpole coefficients have been determined from the average link in Landau gauge: 
$u_{\mu} = 1/3~\langle {\rm tr}~U_\mu(x) \rangle_{\rm Landau}$. 
For brevity we will refer to this as the {\it Landau} scheme.
Any other choice for $(c_s,c_t)$ will have the same
continuum limit, but with this prescription we expect only small
${\cal O}(\alpha_s a)$ discretization errors. 

Our action is a generalization of the ``Fermilab action'' \cite{Fermilab},
which can be considered as a special case of our action with $\xi=1$. It has been demonstrated 
\cite{aniso:klassen-latproc} that the mass dependence of the input parameters 
is much smaller for an anisotropic lattice ( $\xi>1$ ), which makes 
the tuning of the parameters easier in the anisotropic case.

\section{Meson mass measurement}
\label{sec:meson_meas}

We extract meson masses from the time dependence of Euclidean-space two-point 
correlation functions, which is the standard procedure to determine hadron masses from 
lattice calculations. For the hadron source and sink, we use both local 
and extended meson operators constructed from fundamental bilinears in the form of

\be
M(x) = \bar \Psi(x) ~\Gamma_i ~\nabla_j [U] ~\nabla_k[U] ~\Psi(x)~~,
\label{eq:operators}
\ee
here $\Gamma_i$ is one of the 16 Dirac $\Gamma$-matrices and $\nabla_j$ is the symmetric
spatial lattice derivative operator (see Eq. \ref{eq:lat_deriv_op}). To obtain the 
dispersion relation, $E({\bf p})$, we project the meson correlator onto several 
different non-zero momenta by inserting the appropriate phase factors, $\exp{(-i{\bf px})}$, at the sink, and sum the sink over a spatial hyper-plane:
\be
 M(t, {\bf p}) = \sum_{\bf x} M(t,{\bf x}) \exp{(-i{\bf px})},
\label{eq:mom_proj}
\ee
where
\be
{\bf p} = \left(\frac{2\pi}{N_s}\right)~ {\bf n}  \hskip 1cm  \mbox{with~~~} {\bf n} = (0,0,0),(1,0,0),(1,1,0),(2,0,0),(2,2,0), \ldots
\ee

To vary the overlap of a meson operator with the ground state and excited states, we 
also implement a combination of various iterative smearing prescriptions for the 
quark fields and gauge links, which is described below.

For the quark fields, we use box sources to approximate
the spatial distribution of the quark field. More specifically, we set the source to
be unity within a spatial hyper-cube and zero elsewhere. This allows us
to vary the extent in a rather straight forward manner. As such a formulation is not gauge
invariant, we have to fix the gauge, and we choose the Coloumb gauge for convenience. Alternatively, 
we can use a gauge invariant formulation, where repreated action of the gauge 
invariant Laplacian generates a Gaussian-like distribution and provides finer control
of the overlap with different states:
\bea
\label{eq:jacobi}
\Psi(x) &\to& \Psi^{(n)}(x)=(1-\frac{\epsilon}{n} \nabla^2[U])^n \Psi(x)
\eea
This scheme is referred to as Jacobi smearing.

For the gauge field, we apply the well established iterative APE-style link fuzzing algorithm \cite{APEFuzzing}: 
\bea
\label{eq:ape}
X_\mu(x) & = &  {U_\mu}(x) + \alpha * \sum_{\rm staples} S(x,\mu)  \nonumber \\
U_\mu^{\prime}(x) & = &{\cal P}_{SU(3)} ~X_\mu(x) \ , 
\eea
where $\alpha$ is the link fuzzing parameter, $S(x,\mu)$ is one of the six staples around
the link $U_\mu(x)$, and ${\cal P}_{SU(3)}$ is a projection back to $SU(3)$ by requiring
the new link $U_\mu^{\prime}(x)$ maximize:
\be
{\rm{Re\, Tr}} \left ( U^{\prime}_{\mu}(x) X_\mu(x)
\right ) \, .
\label{eq:proj_fuzz}
\ee
The maximization is achieved by several Cabibbo-Marinari 
pseudo-heatbath update steps (fuzzing CM hits).
We repeat the fuzzing step defined by Eq. \ref{eq:ape} for N times (N is the fuzzing 
level). Link fuzzing is quite effective in suppressing gluon excitations and has been widely used in studies of glueball and hybrid hadrons.

This setup allows us to extract reliably both the ground state energies and 
their excitations from correlated multi-state fits to several smeared 
correlators, $C^s(t)$, with the same $J^{PC}$:
\be
C^{s}(t, {\bf p}) \equiv \langle M(t,{\bf p})~M^s(0,{\bf p}) \rangle = \sum_{i=1}^{\rm n_{fit}}~a_i^s({\bf p})~\left(e^{-E_i({\bf p})~t} + e^{-E_i({\bf p}) (N_t -t)} \right)~~.
\label{eq:fit_theory}
\ee
Since we are working in a relativistic setting, the second term takes into account the backward
propagating piece from the temporal boundary. 

We construct meson operators that have definite lattice quantum number $R^{PC}$, in which $R$ is one of 
the five irreducible representations of the hyper-cubic group: $A_1$, $A_2$, $E$, $T_1$, and $T_2$. A brief discussion of the hyper-cubic group is given in appendix \ref{appendix:group},
including our choice of bases that carry the irreducible representations and some relevant 
Clebsch-Gordan coefficients. 
However, determining the continuum quantum number $J^{PC}$ of an operator is complicated
by the fact that the mapping from the finite number of irreducible representations of 
hyper-cubic group to the infinite number of irreducible representations of the continuous 
rotation group is non-unique. Eq. \ref{tab:cubic_to_cont} summarizes the mapping from $R$ to the first few $J$ numbers (up to 4).

\bea
\begin{array}{ll}
	A_1 & \to J=0, 4 \\
	A_2 & \to J=3  \\
	T_1 & \to J=1, 3, 4\\
	T_2 & \to J=2, 3, 4\\
	E   & \to J=2, 4
\end{array}	
\label{tab:cubic_to_cont}
\eea

In most cases, we are only interested in the lowest lying state, which can be extracted from a one-state fit at large time separation
or from a multi-state fit at shorter time separation. However, the lowest state does not always 
correspond to the lowest $J$ number, the nature of a state (orbital excitation,
radial excitation, gluonic excitaion, and etc.) must also be taken into account. A notable example is 
meson operator $b_1 \times D\_T_2$ (detailed in Table \ref{tab:mesonop_complete}), which projects to both $J=2$ and $J=3$ according to table \ref{tab:cubic_to_cont}. The first guess is that the ground state has $J=2$ and is degenerate with the ground state of operator $b_1 \times D\_E$. However, $2^{+-}$ is an exotic hybrid meson , which is much heavier than the non-exotic $3^{+-}$ state. Therefore, the ground state of 
$b_1 \times D\_T_2$ should have quantum number $J=3$ and be degenerate with $b_1 \times D\_A_2$, which is confirmed by our calculation.
Table \ref{tab:mesonop_complete} summarizes all the meson operators used in this work and
the quantum numbers (both lattice and continuum) of the ground states. Our naming convention is explained in appendix \ref{appendix:mesonop}. 

Measuring meson correlators using extended operators is much more expensive than those using 
local operators, because more quark propagators are needed.  We take some steps to improve
the computational efficiency, which are explained in detail in section \ref{sec:simulation}.

\section{Simulation details}
\label{sec:simulation}

The strategy and first results for charmonium have already been presented in 
\cite{aniso:klassen-latproc,aniso:pchen}. The basic idea is to control large lattice spacing artifacts from the heavy quark mass by adjusting the temporal lattice spacing, $a_t$, 
so that $m_q a_t < 1$. We use $\xi=2$ lattices with $\beta=$ 5.7, 5.9, and
6.1. Table \ref{tab:charm_simulations} summarizes the parameters of our simulations.

For the generation of quenched gauge field configurations
we employ a standard heat-bath algorithm as it is also used for
isotropic lattices. The only necessary modification for
simulating an anisotropic plaquette action is to rescale all 
temporal links with the bare anisotropy, $\xi_0$. Depending on the gauge coupling and lattice size, we measure hadron propagators
every 100-400 sweeps in the update process, 
which is sufficiently long for the lattices to decorrelate.

The renormalized anisotropy is related to the bare parameters through:
\be
\xi = \eta(\xi,\beta)~\xi_0  .
\ee
A convenient parametrization for $\eta$ is given in Ref. \cite{aniso:klassen-latproc}
\be
\eta(\xi,\beta) = 1 + (1-\frac{1}{\xi})~\hat \eta_1(\xi)~\frac{1-0.55055~g^2}{1-0.77810~g^2}~\frac{g^2}{6},
\label{eq:renorm_aniso}
\ee
where $\hat \eta_1(\xi)$ is also determined non-perturbatively 
in Ref. \cite{aniso:klassen-latproc} and $\eta(\xi,\infty) = 1 $ as it ought to be.
We use this form to determine the appropriate $\xi_0$
for each value of $\beta=6/g^2=$ 5.7, 5.9, and 6.1 for $\xi=2$.

Although the fine temporal resolution of an anisotropic lattice increases the number of
measurements at short time separations, optimization of the meson operators is also quite 
important. The correlation between measurements of correlation function at 
different time slices reduces the amount of ``useful'' information for a correlated 
fit (in the case of strong correlation between neighboring time slices, little can be gained
by using a higher anisotropy). Using multiple operators (with different optimization parameters) 
and applying multi-state fits can provide stronger constraints for the fits. A source of added difficulty is the possibility that the short distance 
correlators may be 
dominated by excited state signals, which can generate a false plateau, causing 
incorrect identification of an excited state as the
ground state. 
We observe that the hybrid states are rather insensitive to the box size or quark field smearing,
but are highly sensitive to gauge link fuzzing.  Appropriate link fuzzing suppresses excited 
states because fuzzing reduces short-distance fluctuations of the gauge field background.  
This should be particularly important for
hybrid states which are expected to have a non-trivially excited gluonic component. We use multiple link fuzzing with the same
quark smearing for hybrid states. In some cases, we employ fuzzing at the sink only to save 
computational time, tests are done to ensure good coupling to ground states.

Appendix \ref{appendix:mesoncoor} describe in detail the calculation of meson correlators.
First of all, we can reduce the number of quark propagators by 
reusing several basic quark propagators, since all the lattice derivative operators we use 
are linear combinations 
of basic derivative operators: unit($I$), first derivative($\nabla_i(i=x,y,z)$), symmetric second 
derivative ($D_i=s_{ijk}\nabla_j\nabla_k$), asymmetric second derivative($B_i=\epsilon_{ijk}\nabla_j\nabla_k$), and symmetric diagonal second derivative $D^{diag}_i=\nabla_i\nabla_i$. It can be 
shown that we 
can construct all the meson correlation functions using these basic quark propagators.
Furthermore, we can use a linear combination of these basic derivative operators to create a ``white noise'' 
source, which creates more than one quantum number, and rely on the sink operator to project 
to the quantum number of interest. Of course, using such ``white noise'' source saves
computation time at the expense of an increased noise level. We carefully tested various
combinations and chose those combinations that provide the best computational efficiency.

\section{Results and discussions}
\label{sec:results}

We present our charmonium spectrum in Figure \ref{fig:charm_spect}. A clear ordering
of meson states according to orbital angular momentum and gluonic excitation is
shown in Fig. \ref{fig:charm_spect}, as well as the much smaller spin splittings.
The low lying S-wave and P-wave mesons lie below the $D\bar{D}$ threshold, while 
the D-wave and F-wave mesons are above the $D\bar{D}$ threshold. Even higher hybrid 
excitations (with exotic quantum numbers $J^{PC} = 1^{-+}, 0^{+-}, 2^{+-}$) are found 
above the $D^{**}D$ threshold. 
A portion of the spectrum (the low lying $^1S_0$, $^3S_1$, and $^1P_1$) is taken from Columbia group's previous work (Ref.\cite{aniso:pchen}).

The lattice scale is set by the $^1P_1-1S$ splitting 
\footnote{We use $^1P_1$ instead of the spin averaged P-wave meson mass 
$1P=1/9 ({^3P_0} + 3{^3P_1} +5{^3P_2})$ because ${^3P_2}$ requires non-local 
operator which is much nosier and  $^1P_1$ is extremely close
to $1P$ ( $^1P_1 - 1P = 0.86$ MeV $\ll 1P-1S$)}. 
Alternatively, we can use 
the Sommer scale $r_0$ \cite{sommerscale},  which was
employed in Ref. \cite{aniso:pchen}.
Lattice scales from both methods (listed in Table \ref{tab:charm_simulations}) agree well 
(within 2-4\%) for charmonium, which, perhaps is not too suprising since $r_0$ is fixed from a
potential model whose parameters were chosen specifically to reproduce the
experimental charmonium spectrum. However, the discrepancy is rather large for bottomonium. 
We choose the $^1P_1-1S$ scale for this study to be consistent with previous NRQCD 
calculations.

\subsection{Extrapolation to the continuum limit}
\label{sec:cont_extrap}
The continuum limit has to be taken to remove the lattice discretization
errors. This is also a major advantage of relativistic over 
non-relativistic QCD approach. 

The continuum extrapolation of higher excitations is carried out as follows. We first calculate
the dimensionless ratio of splittings between an excited state and the $1S$ state (the spin-averaged mass of S-wave mesons): 
$R(a_s)= (M - M(1S)) / (M(^1P_1)-M(1S)) $ for each lattice spacing $a_s$, and 
then extrapolate the ratio $R(a_s)$ to the continuum limit $a_s=0$. In the end, the continuum 
result of the excited state is obtained by:
\be
 M|_{a_s=0} \ = \ R|_{a_s=0} * (^1P_1-1S)_{exp} + M(1S)_{exp}
\label{eq:mass_cont_extrap}
\ee
where $(^1P_1-1S)_{exp}=458.2$ MeV and $M(1S)_{exp}=3067$ MeV are experimental values.
Because the splitting between an excited state and the $1S$ state is rather insensitive 
to the $1S$ mass (or the quark mass), the imperfect tuning of $1S$ to its experimental value 
has negligible effect on the
physical values obtained from Equation \ref{eq:mass_cont_extrap}. 
However, the incorrect running of the coupling in a quenched
simulation will still cause a  mismatch of scales perceived by excited states and
by the $^1P_1-1S$ splitting.  In contrast, the fine structure splittings
are very sensitive to the quark mass, so the deviation of the $1S$ mass from its experimental value
should be taken into consideration when quoting the physical values for the spin splittings. 

Using tadpole improved coefficients, we are able to remove the leading ${\cal O}(a_s)$ errors,
which has be shown by previous studies \cite{aniso:klassen-latproc,aniso:pchen,aniso:cppacs-charm}.
Therefore, we extrapolate to the continuum limit assuming that the leading lattice spacing error is
${\cal O}(a_s^2)$.

\subsection{Spin splittings}
\label{sec:result:spin}

We first discuss our results for the spin splittings to demonstrate the necessity of a fully 
relativistic treatment, which is crucial for studying the fine structure of charmonium.

The Columbia group has studied in detail this spin splittings  \cite{aniso:pchen},
but the $^3P_2$ state was missing because only local operators were used. Here, we measure
all three P-wave triplet states $^3P_J$ ($J=0,1,2$)  and calculate the P-wave triplet splitting ratio 
$R_{fs}=(^3P_2 - ^3P_1)/(^3P_1 - ^3P_0)$, which is highly sensitive to relativistic corrections but 
less affected by the systematic errors originating from fixing the overall scale in a quenched 
calculation. It has been shown that the velocity ($v / c$) expansion of NRQCD converges poorly for 
charmonium spin splittings \cite{nrqcd:trottier-rel,nrqcd:trottier-spin}. 
Our $R_{fs}$ results are given in table \ref{tab:other}. The result at $\beta=6.1$ has a large 
error because the plateau of the fit sets in at a rather large temporal separation.
Our result for the P-wave triplet splitting ratio $R_{fs}$ is 0.47(13) in the continuum limit,
which agrees well with the experimental value of 0.478(5). A comparison with a NRQCD calculation \cite{thesis:manke} is given in Figure \ref{fig:3pj_sumamry}.
Our result is also consistent with CP-PACS result from anisotropic lattices with anisotropy $\xi=3$ \cite{aniso:cppacs-charm}. For a detailed discussion on the hyperfine splitting, we
refer the reader to Ref. \cite{aniso:pchen}.

\subsection{Excited states}
\label{sec:result:excited}

We first discuss our result for exotic hybrid states, in particular the $1^{-+}$ meson. 
Figure \ref{fig:eff_mass_rhoxB_T1_b61} is an effective mass plot of the exotic hybrid  
meson $1^{-+}$, with each data set corresponding to a particular fuzzing parameter $\alpha$
as defined in Eq. \ref{eq:ape}. With a very fine temporal resolution, we are able to obtain 
many measurements at short
time separations to facilitate multi-state fits and extract both the ground
state and excited state masses more precisely. By employing different quark smearing and 
gauge link fuzzing parameters, we obtain different coupling to the ground and excited states (referred to as different ``channels'' later). 
With multiple channels, the multi-state fits are quite stable. We fit the data to 2-state and in some cases 3-state ansatz to 
obtain the ground state mass. Figure \ref{fig:fit_rhoxB_T1_b61} shows the multi-state
fits for meson $1^{-+}$, the ground state mass obtained from 1-state fit at large time separation
($t > 7$) is consistent with the results from a 2-state fit ($t>4$) and a 3-state fit ($t>1$),
which rules out contamination of the ground state by excited states and clearly demonstrates 
the advantage of anisotropic lattices. For the hybrid states, we did not observe as strong a
dependence  on the quark field smearing as seen for the conventional $\bar{q}q$ mesons, which indicates 
that the excitation is likely to originate from the excitation of gluonic constituents.
For a self-consistency check, we measured the $1^{-+}$ exotic meson with 3 different operators ($a_0^{\prime}\times \nabla$, $b_1 \times \nabla\_T_1$, and $\rho \times B\_T_1$) and obtained consistent values. Fig. \ref{fig:fit_compare_op_extrap_1-+} shows the
continuum extrapolation (${\cal O}(a^2)$) using three different operators, with 
the continuum value being 4.428(41) GeV, 4.409(75) GeV and 4.41(18) GeV 
for $\rho \times B\_T_1$, $a_0^{\prime}\times \nabla$, and $b_1 \times \nabla\_T_1$ irrespectively. The operator $\rho \times B\_T_1$ has the best signal and smallest statistical error, so we quote the final result  4.428(41) GeV for the $1^{-+}$ mass from this operator. 

Various models of QCD, such as the bag models \cite{bagmodel:barnes,bagmodel:charm}, 
the flux-tube model \cite{flux_tube}, and the QCD sum rules \cite{sumrule:1982,sumrule:heavy}  
have been used to study these hybrid states. For the charmonium hybrid meson, the adiabatic bag 
model predicts a $1^{-+}$ $c\bar{c}g$ state of 3.9 GeV \cite{bagmodel:charm} and the QCD 
sum rules predicts 4.1 GeV \cite{sumrule:heavy}. 
We show the comparison of our $1^{-+}$ result to previous lattice results \cite{hybrid-review:page} 
in Fig. \ref{fig:fit_compare_1-+_others}. The result from isotropic lattices 
(MILC \cite{milc_hybrid}) has large systematic uncertainties due to a possible contamination 
from excited states. Previous
results from NRQCD calculations \cite{NRQCDHybrids-ukqcd,CAPACS-manke-hybrid1} are quite 
close to ours, which can be explained by 
the expectation that the quarks in a hybrid meson move more slowly than those in the lowest lying 
$c\bar{c}$ states (as indicated by the hybrid potential study \cite{hybrid-potential-lat}), 
thus making the relativistic corrections small. However, the NRQCD calculations are done 
on rather coarse lattices and the masses can not be extrapolated to the continuum
limit to control the lattice discretization errors.

In addition to the exotic $1^{-+}$ meson, we also measured two other exotic hybrid mesons: 
$0^{+-}$ and $2^{+-}$. They are heavier than the $1^{-+}$ meson, with masses of 4.70(17) GeV and 
4.895(88) GeV, respectively. No clear signal is obtained for $0^{--}$, which is predicted by phenomenological 
models to be even more massive than the $2^{+-}$ state. The QCD Sum rules gives an estimate of
5.9 GeV \cite{sum_rules_hybrid} for the $0^{--}$ state, around 1 Gev above the $2^{+-}$ state.
Clearly it is challenging to extract such a huge excitation from the noise. Higher anisotropies 
may help, but a more efficient operator than our $a_1 \times \nabla\_A_1$ is probably 
also necessary.

Most of our meson operators which can potentially couple to non-exotic hybrid states actually mix strongly 
with the conventional $q\bar{q}$ mesons with the same quantum number $J^{PC}$, as indicated by the
degenerate masses from different operators shown in Table \ref{tab:result}. For example, operator $\pi \times B \_T_1$ has an explicit chromo-magnetic field, making it a potential operator for the hybrid meson $1^{--}$. However, our calculation shows that it has the same mass as the conventional $1^{--}$ 
(see Table \ref{tab:result}). 
The only exception we find is the $2^{-+}$ signal 
from $\rho \times B\_T_2$, which is a very good candidate for a non-exotic hybrid meson. First, its dependence on quark field smearing and gauge link fuzzing is identical to the exotic 
hybrid mesons $1^{-+}$ from $\rho \times B\_T_1$. Second, its mass is much larger than the
conventional $\bar{q}q$ meson. However, further mixing test must be done to rule out the 
possibility that it is the radial excitation of $L=2$ conventional $2^{-+}$.

A selection rule analysis \cite{hybrid-decay:page,hybrid-decay-SS-page} shows that the width of a 
charmonium hybrid meson  is narrow if it lies below the 
$D^{**}D$ threshold (4.287 GeV). Our result for the exotic hybrid $1^{-+}$ is 4.365(47) GeV, slightly above the $D^{**}D$ threshold,
so the hybrid meson is expected to be broad.
However, the conclusion is not final due to the ambiguity of scale setting procedure for 
quenched calculations. 

As a new result from lattice QCD, we also obtain reliable predictions for higher, orbitally 
excited mesons with $L=2$ and 3 (D-wave and F-wave), which provide more information about 
the long range interaction than S-wave or P-wave mesons.
The D-wave mesons ($J^{PC}=
2^{-+}$, $2^{--}$, $3^{--}$) are around 3.9 GeV, above the $D\bar{D}$ threshold (3.7 GeV),
and F-wave mesons ($J^{PC}=3^{++}$, $3^{+-}$) lie at 4.15 GeV, above the $D\bar{D}$ threshold,
but still below the  $D^{**}D$ threshold. 

\subsection{Systematic errors}
\label{sec:result:syserr}

The excited states have larger spacial extent than the lowest lying states, so
the results for excited states are more likely to suffer from finite volume effects
than the lowest lying states. Therefore, we compare the masses of all
excited states obtained from two different lattice sizes ($N_s=8$ and $N_s=16$ at  $\beta=5.7, \xi=2$)
and find almost no finite size effect, as shown in table 
\ref{tab:vol_dep}. This is consistent with previous NRQCD study \cite{CAPACS-manke-hybrid1,Drummond:1999db}, 
which found no finite volume effect for a spacial extent larger than 1.2 fm. 
So we conclude that there is no discernible finite volume effect on our excited state masses.

The second source of error is the imperfect tuning of the bare velocity of light $\nu_t$,
which is found to have a rather big effect on the fine structure \cite{aniso:pchen}. Using
a $\beta=5.7$, $\xi=4$, $8^3 \times 64$ lattice, we vary $\nu_t$ by about 10\%, which 
corresponds to about 5\% change in the renormalized velocity of light. The resultes are
given in table \ref{tab:mass_vt_dep}, 
the changes in excited masses are all less than 2\% and negligible within errors.

The quenched approximation is the largest source of systematic error. The running
of the coupling constant is not correct without the dynamic light quarks. The ratio
$R_{2S}=(2S-1S)/(^1P_1-1S)$ from our quenched lattices  is 1.55(9), which is 15\%
higher than the experimental value of 1.34. It affects our result in two ways.
First, we convert our lattice results to physical units with lattice scales set
by the $^1P_1-1S$ splitting, which is ambiguous due to quenched approximation. Second, we tune
our quark mass to the charmed quark mass by matching the $1S$ mass to its experimental value
of 3.067 GeV using this $^1P_1-1S$ scale, which in turn may affect our final results.
For our excited spectrum, the later can be neglected because the mass splitting 
between a excited state and lowest S-wave meson mass $1S$ is insensitive to quark mass.
To demonstrate this mass insensitiveness, we measure the excited states on a $\beta=5.7$,
$\xi=4$ lattice with two different quark masses $m=0.32$ and $m=0.25$ (corresponding
to $1S$ mass of 3.383 GeV and 3.029 Gev) and find no mass dependence 
(as shown in table \ref{tab:mass_vt_dep}.
However, for the spin splitting which is approximately inversely proportional to the quark
mass, the effect is large (eg. a 5\% correction (increase) in scale will cause 
a 5\% increase in the $1S$ mass, which makes the spin splitting in lattice units 5\% 
smaller than what is should be, so the final result for the spin splitting is 10\% larger).

\section{Conclusion}
\label{sec:conclusion}
We have accurately determined the masses of many new and yet unobserved charmonium
states (both conventional $c\bar{c}$ mesons and hybrid $c\bar{c} g$ mesons) from 
first principles using a fully relativistic anisotropic lattice QCD action.
While all the new states are predicted above the $D\bar{D}$ threshold, we believe
that the exotic hybrids may still be sufficiently long-lived to warrant an experimental
search in the energy region near and above the $D^{**}D$ threshold of 4.287 GeV.
This will be an important effort to validate non-perturbative QCD at low energies.
We add further evidence that a fully relativistic treatment of the heavy quark system is
well suited to control the large systematic errors of NRQCD. In addition, the anisotropic
lattice formulation is a very efficient framework for calculations requiring high temporal
resolutions, which is crucial for the study of high excitations.
The quenched approximation, which is the major remaining systematic error in our calculation, 
will be addressed by future full QCD calculation. Full QCD 
anisotropic lattice action has been implemented for staggered fermion and used for 
thermodynamics study \cite{ludmila-aniso}. In addition to the heavy-heavy system, 
an anisotropic lattice is also a natural framework to study heavy-light systems.

\section*{Acknowledgments}
We would like to thank Prof. Norman Christ for his constant support and many valuable. 
suggestions. We are grateful for many stimulating discussions with to Prof. Robert 
Mawhinney. 
This work is supported by the U.S. Department of Energy. Numerical simulations 
were conducted on the QCDSP super computers at Columbia University and Brookhaven 
RIKEN-BNL Research Center.


\appendix

%
\section{Conventions}
\label{appendix:conventions}
In this section, we describe the conventions used in this paper.
The variable $x$ specifies the coordinates in the 4-dimensional (space-time) 
volume, with extent $N_s$ along the spacial directions and $N_t$ along
the time direction. The spacial physical lattice size are given by  
$L_s=N_s * a_s$. ${\bf x}$ is the 3-dimensional spacial coordinate, and 
$t$ is the time coordinate. $\Psi(x)$ or $q(x)$ represents a fermion field.

The definitions of $s_{ijk}$ and $S_{\alpha jk}$ ($\alpha$=1,2) are:
\be
	s_{ijk} = |\epsilon_{ijk}|
\label{eq:sijk}
\ee
\be
	S_{\alpha jk}=0 (j \ne k), S_{111}=1, S_{122}=-1, S_{222}=1, S_{233}=-1
\label{eq:Sijk}
 \ee

%
%
\section{Meson correlator and meson operators}
\subsection{Meson correlator calculation}
\label{appendix:mesoncoor}
We describe how to construct meson correlators from quark propagators.  For simplicity, 
we discuss the correlation function between sink meson operator ${\cal M}(t)=\bar\Psi_2({\bf x_2},t) \stackrel{\rightarrow}{O_i} \Gamma_{\alpha} \Psi_1({\bf x_1}, t)$, 
and source operator $M(t_0)=\bar\Psi_2({\bf y_2}, t_0) \stackrel{\rightarrow}{O_j} \Gamma_{\beta} \Psi_1({\bf y_1}, t_0)$, 
in which $\Gamma_{\alpha}$ and $\Gamma_{\beta}$ are a combination of Dirac $\gamma$-matrices, and $\stackrel{\rightarrow}{O}_i$ and $\stackrel{\rightarrow}{O}_j$ are lattice derivative operators. More generally, 
the source operator can be any operator (or a combination of different meson
operators) that creates the state of interest.

The correlator is calculated as follows:

  \bea
  \label{eq:corr_calc}
	\langle {\cal M}^{\dag}(t)M(t_0)\rangle & = &
	\langle \Psi_1^{\dag}({\bf x_1}, t) \Gamma_{\alpha}^{\dag} \stackrel{\leftarrow}{O}_i^{\dag} \g_4 \Psi_2({\bf x_2}, t) ~~~
	\bar\Psi_2({\bf y_2}, t_0) \stackrel{\rightarrow}{O}_j \Gamma_{\beta} \Psi_1({\bf y_1}, t_0) \rangle \nonumber \\
	&=& \langle \bar\Psi_1({\bf x_1}, t) \g_4 \Gamma_{\alpha}^{\dag} \g_4 \stackrel{\leftarrow}{O}_i^{\dag} ~~\Psi_2({\bf x_2}, t)
	\bar\Psi_2({\bf y_2}, t_0)~~ \stackrel{\rightarrow}{O}_j \Gamma_{\beta} \Psi_1({\bf y_1}, t_0) \rangle \nonumber \\
	&=& \langle \g_4 \Gamma_{\alpha}^{\dag} \g_4 ~~G_2({\bf x_2}, t; {\bf y_2}, t_0)~~\Gamma_{\beta} ~~
	G({\bf y_1}, t_0 ; {\bf x_1}, t) \stackrel{\leftarrow}{O}_i^{\dag} \rangle \nonumber \\
	&=& \langle \g_5 \g_4 \Gamma_{\alpha}^{\dag} \g_4 ~~ G_2({\bf x_2}, t; {\bf y_2}, t_0)~~\Gamma_{\beta} \g_5 ~~[ \stackrel{\leftarrow}{O}_i G({\bf x_1}, t ; {\bf y_1}, t_0)]^{\dag} \rangle \nonumber \\
	&=& \langle \Gamma_{\alpha}^{\prime} ~~G_2({\bf x_2}, t; {\bf y_2}, t_0)~~ \Gamma_{\beta}^{\prime} 
	~~[ \stackrel{\leftarrow}{O}_i G({\bf x_1}, t ; {\bf y_1}, t_0)]^{\dag} \rangle
  \eea

Here, we first set up the source (box or smeared) and apply the derivative operator
$\stackrel{\rightarrow}{O}_j$, then we calculate the quark propagator $G({\bf y_1}, t_0 ; {\bf x_1}, t)=\langle \bar\Psi_1({\bf x_1}, t) \stackrel{\rightarrow}{O}_j 
\Psi_1({\bf y_1}, t_0) \rangle$. Utilizing $G({\bf y_1}, t_0; {\bf x_1}, t) = \g_5 G^{\dag}({\bf x_1}, t;{\bf y_1}, t_0) \g_5$, we arrive at equation \ref{eq:corr_calc}.
Then we apply the sink derivative operator $\stackrel{\rightarrow}{O}_i$, insert gamma matrices 
$\Gamma_{\alpha}^{\prime}=\g_5 \g_4 \Gamma_{\alpha}^{\dag} \g_4$ and $\Gamma_{\beta}^{\prime}=\Gamma_{\beta} \g_5$, 
and do color and Dirac indices contraction. 
We need two quark propagators: $G({\bf x_1}, t; {\bf y_1}, t_0)$ and $G_2({\bf x_2}, t; {\bf y_2}, t_0)$. In practice, we can reuse the same quark propagators for multiple meson operators.

%
%
\subsection{Meson operators and naming convention}
\label{appendix:mesonop}
Meson operators with no derivative operators are named $a_0$, $a_0^\prime$ , $\pi$, $\rho$, 
$a_1$, and  $b_1$, as listed in the second row of table \ref{tab:mesonop_brief}. 
A meson operator with a lattice derivative operator is named $X \times Y\_R$, where X 
represents the gamma matrix, 
Y is the derivative operator ($\nabla$, D or B), and R indicates the $O_h$ irreducible
representation of the meson operator (R is usually omitted if it is unique).

%
%
\section{Hyper-cubic group}
\label{appendix:group}

The five irreducible representations of hyper-cubic group $O_h$ and our choice of basis vectors for 
these representations are:

\bea
A_1:  & \psi^1 = &(x^2+y^2+z^2) \nonumber\\
A_2:  &  \psi^2 = & xyz \nonumber\\
E:    & (\psi^3_1, \psi^3_2)=&(x^2-y^2, y^2-z^2)\\
T_1:  & (\psi^4_x, \psi^4_y, \psi^4_z)= &(x, y, z) \nonumber \\
T_2:  & (\psi^5_{yz}, \psi^5_{xz}, \psi^5_{xy})=& (yz+zy, zx+xz, xy+yx) \nonumber
\eea

The decomposition of often used direct products of hyper-cubic group irreducible representations:

\bea
\begin{array}{ccc}
	T_1 \bigotimes T_1 & = & A_1 \bigoplus T_1 \bigoplus T_2 \bigoplus E\\
	T_1 \bigotimes T_2 & = & A_2 \bigoplus T_1 \bigoplus T_2 \bigoplus E\\
\end{array}
\eea

Clebsch-Gordon coefficients are defined as:\\
\bea
\psi(J, M) = \sum_{m_1,m_2} C(J,M | J1, m1; J2, m2)\ u(J1,m1)\ v (J2, m2)
\eea

Table \ref{tab:CG_T1T1},  \ref{tab:CG_T1T2},  
are Clebsch-Gordon coefficients used to construct the meson operators in table \ref{tab:mesonop_complete}

\input runs

\input results_all
\input tables

\input plots_all

\end{document}

%% file: runs.tex
\begin{table}[htb]
\begin{center}
\begin{tabular}{l|lllll}
$(\beta,\xi)$                  &  (5.7, 2) & (5.9, 2)   &  (6.1, 2)	 & (5.7, 2) \\
$(N_s,N_t)$                    &  (8, 32)  & (16, 64)   &  (16, 64)  	 & (16, 32)\\
configs                        &  1950     &  1080      &  1010      	 & 1200    \\
separation                     &  100      &  200       &  400      	 & 100    \\
\hline                                                               
$a_t^{-1}$ ($r_0$) [GeV]                & 1.905(5)     & 2.913(9)     & 4.110(22) & 1.905(5) \\
$a_t^{-1}$ ($^1P_1-1S$) [GeV]                & 1.945(26)     & 3.021(34)     & 4.292(49) & 1.924(25)  \\
$a_s$ [fm]                      & 0.1974(26) & 	0.1307(15) &	0.0920(11) & 0.2052(27) \\
$L_s$ [fm]                      & 1.624     & 2.091	 & 1.472 	& 2.283 \\
\hline                                                                
$\xi_0$                         & 1.654729 & 1.690713  & 1.718306  	& 1.654729  \\
$u_{0s}$                         & 0.7762   & 0.8091   & 0.8280  	& 0.7762      \\
$u_{0t}$                         & 0.9394   & 0.9504   & 0.9569  	& 0.9394   \\
$u_{0t}/u_{0s}$                  & 1.2103   & 1.1746   & 1.1557  	& 1.2103   \\
$\xi/\xi_0$                     & 1.208657 & 1.1829329  & 1.163937 	& 1.208657  \\
$a_tm_q$                        & 0.51     & 0.195     	& 0.05    	& 0.51	   \\
($\nu_s, \nu_t$)                & (1, 1.01)& (1, 1.09)  & (1, 1.12)	&   (1, 1.01) \\
$c_s$                           & 2.138    & 1.889   & 1.7614  		& 2.138  \\
$c_t$                           & 1.3252   & 1.2055   & 1.1431 		& 1.3252 \\
$c(0)$                          & 1.000(2) & 0.984(1) & 0.984(3) 	& 0.991(3) \\
\hline                                                                          
Box size 			& 7	   &  16          &   16     &  7      \\
Fuzzing $\alpha$		& (1/30, 1/6, 2/3) &(1/30, 1/6, 1) &  (1/30, 1/6, 1) & (1/30, 1/6, 2/3)  \\
Fuzzing Level			& 5	   & 6		& 7	& 5	   \\
Fuzzing CM hits 			&  8        &	6	& 6	& 8	     \\
\end{tabular}
\vskip0.2cm
\caption{Charmonium spectrum simulation parameters. 
}
\label{tab:charm_simulations}
\end{center}
\end{table}

%% file: results_all.tex
\begin{table}[ptb]
\begin{center}
\begin{tabular}{|c|c|c|c|c|c|c|}
($\beta, \xi$) &          & (5.7, 2)        & (5.9, 2)       & (6.1, 2)    & continuum(${\cal O}(a^2)$) & continuum(${\cal O}(a)$)\\
\hline
   Operator      & $J^{PC}$ & $E_0$ (GeV) & $E_0$ (GeV)  & $E_0$ (GeV)  & $E_0$ (GeV) & $E_0$ (GeV) \\
\hline
       $a_1\times B\_A_1$    	 &  $0^{+-}$			 & 4.525(51) & 4.63(15)  & 4.82(17) & 4.70(17)   & 4.84(27)  \\
\hline
  $a_0^\prime \times \nabla$		 &  $1^{-+}$		 & 4.30(14) & 4.406(64) & 4.486(52) & 4.409(75)   & 4.52(15)  \\
\hline
	$b_1\times\nabla \_T_1$    	 &  $1^{-+}$		 & 4.34(16) & 4.45(10) & 4.52(22)  & 4.41(18)   & 4.51(34)  \\
\hline
	$\rho \times B\_T_1$   	 &  $1^{-+}$		 & 4.439(16) & 4.524(27) & 4.507(56)   & 4.428(41)   & 4.465(68)  \\
\hline
  $a_0^\prime \times D$		 &  $2^{+-}$		 & 4.494(97)   & 4.717(54)   & 4.95(12)  & 4.82(10)   & 5.09(19)  \\
\hline
       $a_1 \times B\_T_2$    	 &  $2^{+-}$		 & 4.710(60)  & 4.868(54)   & 5.03(11)   & 4.895(88)   & 5.06(16)  \\
\hline
	$b_1 \times D\_E$	    	 &  $2^{+-}$			 & 4.51(14)  & 4.674(73)  & 4.88(12)  & 4.76(12)   & 4.98(23)  \\
\hline
	$\pi$				 &   $0^{-+}$		 & 3.0870(20)   & 3.1245(10)  & 3.1249(21)   & 3.0082(19)   & 3.0099(38)  \\
\hline
	$a_0$				 &   $0^{++}$	  	 & 3.5547(58)  & 3.5944(31)   & 3.5730(42)   & 3.4628(48)   & 3.4507(91)  \\
\hline
	$\rho \times\nabla \_A_1$   	 &  $0^{++}$		 & 3.4965(27)  & 3.5466(91) & 3.530(12)  & 3.426(11)   & 3.430(18)  \\
\hline
       $a_0\times \nabla$			 &  $1^{--}$		 & 3.1611(19)   & 3.2033(12)  & 3.1958(21)   & 3.0852(20)   & 3.0856(38)  \\
\hline
	$\pi \times B\_T_1$  	 &  $1^{--}$			 & 3.1640(27)  & 3.20430(60)  & 3.1992(52)  & 3.0864(23)   & 3.0867(53)  \\
\hline
	$\rho$				 &   $1^{--}$		 & 3.1646(20) & 3.2053(10)  & 3.1980(21)  & 3.0866(19)   & 3.0860(38)  \\
\hline
	$\rho \times D\_T_1$   	 &  $1^{--}$		 & 3.157(53)   & 3.2046(45)  & 3.297(43)   & 3.129(37)   & 3.208(84)  \\
\hline
       $a_1 \times B\_T_1$    	 &  $1^{+-}$		 & 3.575(21)   & 3.641(26)   & 3.640(26)  & 3.538(29)   & 3.558(48)  \\
\hline
	$b_1$				 &   $1^{+-}$		 & 3.6006(51) & 3.6483(52)   & 3.653(13)   & 3.5388(87)   & 3.549(15)  \\
\hline
	$b_1 \times D\_T_1$    	 &  $1^{+-}$		 & 3.5755(39) & 3.6200(91)  & 3.610(17)   & 3.503(13)   & 3.505(21)  \\
\hline
	$\pi \times \nabla$ 	    	 &  $1^{+-}$			 & 3.562(16)  & 3.605(12)  & 3.602(17)  & 3.490(17)   & 3.493(31)  \\
\hline
	$a_1$				 &   $1^{++}$	 	 & 3.5975(61)   & 3.6504(21)   & 3.6262(63)  & 3.5324(52)   & 3.531(11)  \\
\hline
       $a_1 \times D\_T_1$    	 &  $1^{++}$		 & 3.5891(78) & 3.602(15) & 3.6161(82)   & 3.497(10)   & 3.492(16)  \\
\hline
	$\rho \times\nabla \_T_1$   	 &  $1^{++}$		 & 3.5362(27)   & 3.5986(91) & 3.580(11)  & 3.482(11)   & 3.493(17)  \\
\hline
       $a_1 \times \nabla \_E$	    	 &  $2^{--}$		 & 3.832(23)   & 3.917(14)  & 3.893(17)   & 3.803(20)   & 3.816(37)  \\
\hline
	$a_1 \times \nabla \_T_2$    	 &  $2^{--}$		 & 3.891(31)  & 3.910(15)  & 3.893(13)   & 3.772(18)   & 3.753(36)  \\
\hline
	$\rho \times D\_T_2$   	 &  $2^{--}$		 & 3.8916(93)   & 3.943(18) & 3.923(13) & 3.815(15)   & 3.814(24)  \\
\hline
	$\pi \times D\_T_2$  	 &  $2^{-+}$			 & 3.8842(93)   & 3.931(14)   & 3.923(13)   & 3.814(14)   & 3.817(23)  \\
\hline
	$\rho \times B\_T_2$   	 &  $2^{-+}$		 & 4.548(21)   & 4.660(33)   & 4.623(64)   & 4.570(50)   & 4.623(82)  \\
\hline
       $a_1 \times D\_E$	    	 &  $2^{++}$			 & 3.6062(39)   & 3.6361(73)   & 3.6427(73)  & 3.5237(78)   & 3.523(12)  \\
\hline
       $a_1 \times D\_T_2$    	 &  $2^{++}$		 & 3.6033(84)   & 3.6324(76)   & 3.6367(60)   & 3.5202(74)   & 3.520(13)  \\
\hline
	$\rho \times\nabla \_T_2$   	 &  $2^{++}$		 & 3.5570(62)  & 3.6207(91)   & 3.605(12)   & 3.508(12)   & 3.521(19)  \\
\hline
	$\rho \times D\_A_2$   	 &  $3^{--}$		 & 3.912(12)   & 3.9760(91)   & 3.958(17)  & 3.865(15)   & 3.879(27)  \\
\hline
	$b_1 \times D\_A_2$    	 &  $3^{+-}$		 & 4.153(29)   & 4.333(39)   & 4.271(47)   & 4.239(48)   & 4.312(79)  \\
\hline
	$b_1 \times D\_T_2$    	 &  $3^{+-}$		 & 4.157(27)  & 4.339(42)   & 4.383(39)  & 4.326(43)   & 4.450(70)  \\
\hline
       $a_1 \times D\_A_2$    	 &  $3^{++}$		 & 4.151(56)   & 4.248(54)   & 4.344(99)   & 4.215(84)   & 4.31(15)  \\
\end{tabular}
\end{center}
\vskip0.2cm
\caption{ 
Charomonium spectrum results expressed in GeV. The meson operators are sorted by the continuum quantum number 
($J^{PC}$) of the ground state. Only the ground state masses are listed in this table. The lattice
scale is set by the $^1P_1-1S$ splitting. Continuum results from ${\cal O} (a_s^2)$ and ${\cal O} (a_s)$ extrapolations are listed in the last two columns.  
The continuum extrapolation procedure is described in section 
\ref{sec:cont_extrap}. The errors quoted here do not include the uncertainties in the physical
scale. 
}
\label{tab:result}
\end{table}

\begin{table}
\begin{center}
\begin{tabular}{llllllllllll}
$(\beta,\xi)$                   &        (5.7, 2)        &      (5.9 ,2)      &  (6.1, 2)  & cont. ${\cal O}(a^2)$ & cont. ${\cal O}(a)$ \\
\hline
$R_{fs}$    &        0.474(50)       &      0.46(8)      & 0.52(20) & 0.47(13) & 0.47(21)\\
\hline
$R_{2S}$    &        1.50(12)        &      1.563(45)    & 1.517(77) &    1.55(9) & 1.54(18)\\ 
\end{tabular}
\vskip0.3cm
\caption{ P-wave spin splitting ratio $R_{fs}=(^3P_2-^3P_1)/(^3P_1-^3P_0)$ and 
          excited state mass ratios ($R_x=(X-1S)/(^1P_1-1S)$).
}
\label{tab:other}
\end{center}
\end{table}

\begin{table}
\begin{center}
\begin{tabular}{llll}
$(\beta,\xi)$                   &        (5.7, 2)        &      (5.7 ,2)     \\
$N_s$			&	8		&	16	\\
$L_s$ [fm]		& 1.624 	&  3.283 \\
\hline
$2^{-+}$   & 0.739((9)       & 0.758(13)\\
$3^{++}$   & 1.006(56)       & 0.992(31) \\
$1^{-+}$   	& 1.294(16)	& 1.281(15)		\\
$0^{+-}$   	& 1.351(51)	& 1.337(37)		\\
$2^{+-}$   	& 1.564(63)	& 1.575(71)		\\
\end{tabular}
\vskip0.3cm
\caption{ Finite volume effects of excited charmonium spectrum.  The values quoted are masses relative
to the $1S$ state.
	}
\label{tab:vol_dep}
\end{center}
\end{table}

\begin{table}
\begin{center}
\begin{tabular}{llll}
$(\beta,\xi)$                   &        (5.7, 4)        &      (5.7 ,4) & (5.7, 4)     \\
$m_q$   & 0.32 & {\bf 0.25} & 0.32 \\
$\nu_t$ & 1.37 & 1.37 & {\bf 1.25} \\
\hline
$2^{-+}$   & 0.801(14)       & 0.807(15) & 0.801(12)\\
$3^{++}$   & 1.062(23)       & 1.078(43) & 1.071(46)\\
$1^{-+}$   	& 1.316(31)	& 1.312(31)	& 1.318(68) 	\\
$0^{+-}$   	& 1.461(46)	& 1.475(72)	& 1.496(57)	\\
$2^{+-}$   	& 1.692(46)	& 1.674(61)	& 1.715(63)	\\
\end{tabular}
\vskip0.3cm
\caption{ Mass and velocity of light dependence of excited charmonium spectrum. 
          The values quoted are masses relative to the $1S$ state. 
	}
\label{tab:mass_vt_dep}
\end{center}
\end{table}

%% file: tables.tex
\begin{center}
\begin{table}[htb]
\begin{tabular}{|c|c|c|c|c|c|c|}
 $O \times \Gamma$ & 1	& $\g_4$ & $\g_5$ & $(\g_i,\g_4\g_i)$ & $\g_5\g_i$ & $(\g_4\g_5\g_i,\g_i\g_j)$ \\
\hline
 1   & $a_0=0^{++}$ & $a_0^\prime={\bf 0}^{+-}$ & $\pi = 0^{-+}$ & $\rho=1^{--}$ & $a_1=1^{++}$ & $b_1=1^{+-}$ \\
\hline
$\nabla_i$ & $1^{--}$  & ${\bf 1}^{-+}$ & $1^{+-}$ & $(0,1,2)^{++}$ & $({\bf 0},1,2)^{--}$ & $(0,{\bf 1},2)^{-+}$ \\
\hline
$B_i$ & $1^{+-}$ & $1^{++}$ & $ 1^{--}$ & $ (0,{\bf 1},2)^{-+}$ & $({\bf 0},1,{\bf 2})^{+-}$ & $(0,1,2)^{++}$\\
\hline
$D_i$ & $2^{++}$ & ${\bf 2}^{+-}$ & $ 2^{-+}$ & $ (1,2,3)^{--}$ & $ (1,2,3)^{++}$ & $ (1,{\bf 2},3)^{+-}$\\
\end{tabular}
\vskip0.3cm
\caption{A quick reference of 
$q\bar q$ and $q\bar q g$ states (exotics in bold font) accessible with $O \times \Gamma$ construction of
meson operator, O is a derivative operator and $\Gamma$ is a combination of Dirac $\gamma$-matrices.
Here $\nabla_i$ is the first derivative operator, $D_i=s_{ijk}\nabla_j\nabla_k$ (i=1,2,3), and $B_i=\epsilon_{ijk}\nabla_j\nabla_k$, where $s_{ijk}$ is the symmetric tensor defined by Eq. \ref{eq:sijk}.
}
\label{tab:mesonop_brief}
\end{table}
\end{center}

\begin{table}[htb]
\begin{tabular}{|l|l|l|l|l|}
Operator             & $O_h$ rep.         & lowest $J^{PC}$ & name & remark \\
\hline
1                      &  $A_1$  & $0^{++}$        & $a_0$ & ${^3}P_0 (\chi_{c0})$  \\
$\g_5$          	&  $A_1$  & $0^{-+}$        & $\pi$ & ${^1}S_0 (\eta_{c})$  \\
$\g_i$           	&  $T_1$  & $1^{--}$        & $\rho$ & ${^3}S_1 (J/\psi)$  \\
$\g_5\g_i$              & $T_1$  & $1^{++}$        & $a_1$ & ${^3}P_1 (\chi_{c1})$\\
$\g_i\g_j$  		& $T_1$ & $1^{+-}$        & $b_1$  & ${^1}P_1 (h_c)$ \\
\hline
$\g_5\nabla_i$               & $T_1$                 & $1^{+-}$      & $\pi \times \nabla$ & \\
$\nabla_i$			 & $T_1$		     & $1^{--}$	     & $a_0 \times \nabla$   &	\\
$\g_4\nabla_i$               & $T_1$                  & $1^{-+}$      & $a_0^{\prime} \times \nabla$ & \\
$\g_i\nabla_i$                 & $A_1$ 	& $0^{++}$      & $\rho \times \nabla\_A_1$  & ${^3}P_0 (\chi_{c0})$ \\
$\e_{ijk}\g_j\nabla_k$         & E	& $1^{++}$      & $\rho \times \nabla\_T_1$     & ${^3}P_1 (\chi_{c1})$ \\
$s_{ijk}\g_j\nabla_k$         & $T_2$	&$2^{++}$      & $\rho \times \nabla \_T_2$     & ${^3}P_2 (\chi_{c2})$ \\
$\g_5\g_i\nabla_i$            & $A_1$	&$0^{--}$      & $a_1 \times \nabla\_A_1$   & exotic \\
$\g_5 s_{ijk}\g_j\nabla_k$    & $T_2$	&$2^{--}$      & $a_1 \times \nabla\_T_2$           & \\
$\g_5 S_{\alpha jk}\g_j\nabla_k$  & $T_2$	&$2^{--}$      & $a_1 \times \nabla\_E$           & \\
$\g_4\g_5\e_{ijk}\g_j\nabla_k$ &  $T_1$	&$1^{-+}$      & $b_1 \times \nabla\_T_1$          & exotic \\
$\g_4s_{ijk}\nabla_j\nabla_k$     & $T_2$	&$2^{+-}$      & $a_0^{\prime} \times D$  & exotic \\
$\g_5\g_iD_i$ 		 & $A_2$  &$3^{++}$	& $a_1 \times D\_A_2$ &	\\
$\g_5 S_{\alpha jk}\g_jD_k$   & E  &$2^{++}$	& $a_1 \times D\_E$ &	\\
$\g_5s_{ijk}\g_jD_k$  	  	  & $T_1$  &$1^{++}$	& $a_1 \times D\_T_1$ &	\\
$\g_5\e_{ijk}\g_jD_k$  	    & $T_2$  &$2^{++}$	& $a_1 \times D\_T_2$ &	\\
$\g_4\g_5s_{ijk}\g_i\nabla_j\nabla_k$  & $A_2$  &$3^{+-}$	& $b_1 \times D\_A_2$ &	\\
$\g_4\g_5 S_{\alpha jk}\g_jD_k$	  & E  &$2^{+-}$	& $b_1 \times D\_E$ &	\\
$\g_4\g_5s_{ijk}\g_jD_k$  	 	  & $T_1$  &$1^{+-}$	& $b_1 \times D\_T_1$ &	\\
$\g_4\g_5\e_{ijk}\g_jD_k$  	  & $T_2$  &$3^{+-}$	& $b_1 \times D\_T_2$ &	\\
$\g_iD_i$ 		 	& $A_2$  &$3^{--}$	& $\rho \times D\_A_2$ &	\\
$s_{ijk}\g_jD_k$  	 	& $T_1$  &$1^{--}$	& $\rho \times D\_T_1$ &	\\
$\e_{ijk}\g_jD_k$  	  	& $T_2$  &$2^{--}$	& $\rho \times D\_T_2$ &	\\
$\g_4\g_5s_{ijk}\nabla_j\nabla_k$    & $T_2$  &$2^{-+}$	& $\pi \times D\_T_2$ &	\\
$\g_5B_i$		  & $T_1$    &$1^{--}$	& $\pi \times B\_T_1$ & \\
$\e_{ijk}\g_jB_k$                & $T_1$	&$1^{-+}$      & $\rho \times B\_T_1$           & exotic \\
$s_{ijk}\g_jB_k$             & $T_2$	&$2^{-+}$      & $\rho \times B\_T_2$           &  \\
$\g_5\g_iB_i$                & $A_1$	&$0^{+-}$      & $a_1 \times B\_A_1$   & exotic \\
$\g_5\e_{ijk}\g_jB_k$      & $T_1$	&$1^{+-}$      &  $a_1 \times B\_T_1$          & \\
$\g_5s_{ijk}\g_jB_k$        & $T_2$	&$2^{+-}$      &  $a_1 \times B\_T_2$          & exotic \\
\end{tabular}
\vskip0.3cm
\caption{Meson operators, names and quantum numbers.
	The quantities $s_{ijk}$ and $S_{\alpha jk}$ are defined by equations \ref{eq:sijk} and \ref{eq:Sijk}.
	}
\label{tab:mesonop_complete}
\end{table}


\begin{table}[h]
\begin{center}
\begin{tabular}{|c||c|c|c|c|c|c|c|c|c|}
        & $u^4_x\ v^4_x$ & $u^4_x\ v^4_y$ &$u^4_x\ v^4_z$ & $u^4_y\ v^4_x$ & $u^4_y\ v^4_y$ & $u^4_y\ v^4_z$ & $u^4_z\ v^4_x$ & $u^4_z\ v^4_y$ &  $u^4_z\ v^4_z$ \\
\hline
\hline
$\psi^1$   & $1/\sqrt{3}$ & 0 & 0 & 0 & $1/\sqrt{3}$ & 0        & 0 & 0 & $1/\sqrt{3}$ \\
\hline
$\psi^3_1$ & $1/\sqrt{2}$ & 0 & 0 & 0 & $-1/2$ & 0      & 0 & 0 & $-1/2$\\
$\psi^3_2$ & $1/2$        & 0 & 0 & 0 & $1/2$ & 0       & 0 & 0 & $-1/\sqrt{2}$\\
\hline
$\psi^4_x$ & 0           & 0 & 0 & 0 & 0 & $1/\sqrt{2}$ & 0     & -$1/\sqrt{2}$ & 0\\
$\psi^4_y$ & 0  & 0 &  -$1/\sqrt{2}$  & 0 & 0   & 0 & $1/\sqrt{2}$ & 0 &0\\
$\psi^4_z$ & 0 & $1/\sqrt{2}$ & 0 & -$1/\sqrt{2}$ & 0 & 0       & 0 & 0 & 0 \\
\hline
$\psi^5_{yz}$& & 0 & 0 & 0 & 0 & $1/\sqrt{2}$   & 0 & $1/\sqrt{2}$ & 0\\
$\psi^5_{xz}$& 0 & 0 &  $1/\sqrt{2}$& 0 & 0 & 0 & $1/\sqrt{2}$ & 0 & 0\\
$\psi^5_{xy}$& 0 & $1/\sqrt{2}$ & 0 &  $1/\sqrt{2}$& 0 & 0      & 0 & 0 & 0\\
\end{tabular}
\vskip0.3cm
\caption{ Clebsch-Gordon coefficients of $T_1 \bigotimes T_1 = A_1 \bigoplus E \bigoplus T_1 \bigoplus T_2 $ }
\label{tab:CG_T1T1}
\end{center}
\end{table}

\begin{table}[h]
\begin{center}
\begin{tabular}{|c||l|c|c|c|c|c|c|c|c|}
        & $u^4_x\ v^5_{yz}$ & $u^4_x\ v^5_{xz}$ &$u^4_x\ v^4_{xy}$ & $u^4_y\ v^5_{yz}$ & $u^4_y\ v^5_{xz}$ & $u^4_y\ v^5_{xy}$ & $u^4_z\ v^5_{yz}$ & $u^4_z\ v^5_{xz}$ &  $u^4_z\ v^5_{xy}$ \\
\hline
\hline
$\psi^2$   & $1/\sqrt{3}$ & 0 & 0 & 0 & $1/\sqrt{3}$ & 0        & 0 & 0 & $1/\sqrt{3}$ \\
\hline
$\psi^3_1$ & 0 & 0 & 0 & 0 & $-1/\sqrt{2}$ & 0  & 0 & 0 & $1/\sqrt{2}$\\
$\psi^3_2$ & $1/\sqrt{2}$         & 0 & 0 & 0 & $-1/\sqrt{2}$ & 0       & 0 & 0 & 0\\
\hline
$\psi^4_x$ & 0           & 0 & 0 & 0 & 0 & $1/\sqrt{2}$ & 0     & $1/\sqrt{2}$ & 0\\
$\psi^4_y$ & 0  & 0 &  $1/\sqrt{2}$  & 0 & 0    & 0 & $1/\sqrt{2}$ & 0 &0\\
$\psi^4_z$ & 0 & $1/\sqrt{2}$ & 0 & $1/\sqrt{2}$ & 0 & 0        & 0 & 0 & 0 \\
\hline
$\psi^5_{yz}$& & 0 & 0 & 0 & 0 & $1/\sqrt{2}$   & 0 & -$1/\sqrt{2}$ & 0\\
$\psi^5_{xz}$& 0 & 0 &  -$1/\sqrt{2}$& 0 & 0 & 0        & $1/\sqrt{2}$ & 0 & 0\\
$\psi^5_{xy}$& 0 & $1/\sqrt{2}$ & 0 &  -$1/\sqrt{2}$& 0 & 0     & 0 & 0 & 0\\
\end{tabular}
\vskip0.3cm
\caption{Clebsch-Gordon coefficients of $T_1 \bigotimes T_2 = A_2 \bigoplus E \bigoplus T_1 \bigoplus T_2 $}
\label{tab:CG_T1T2}
\end{center}
\end{table}

%% file: plots_all.tex
\begin{figure}[htb]
\begin{center}
\vskip0.5cm
\epsfig{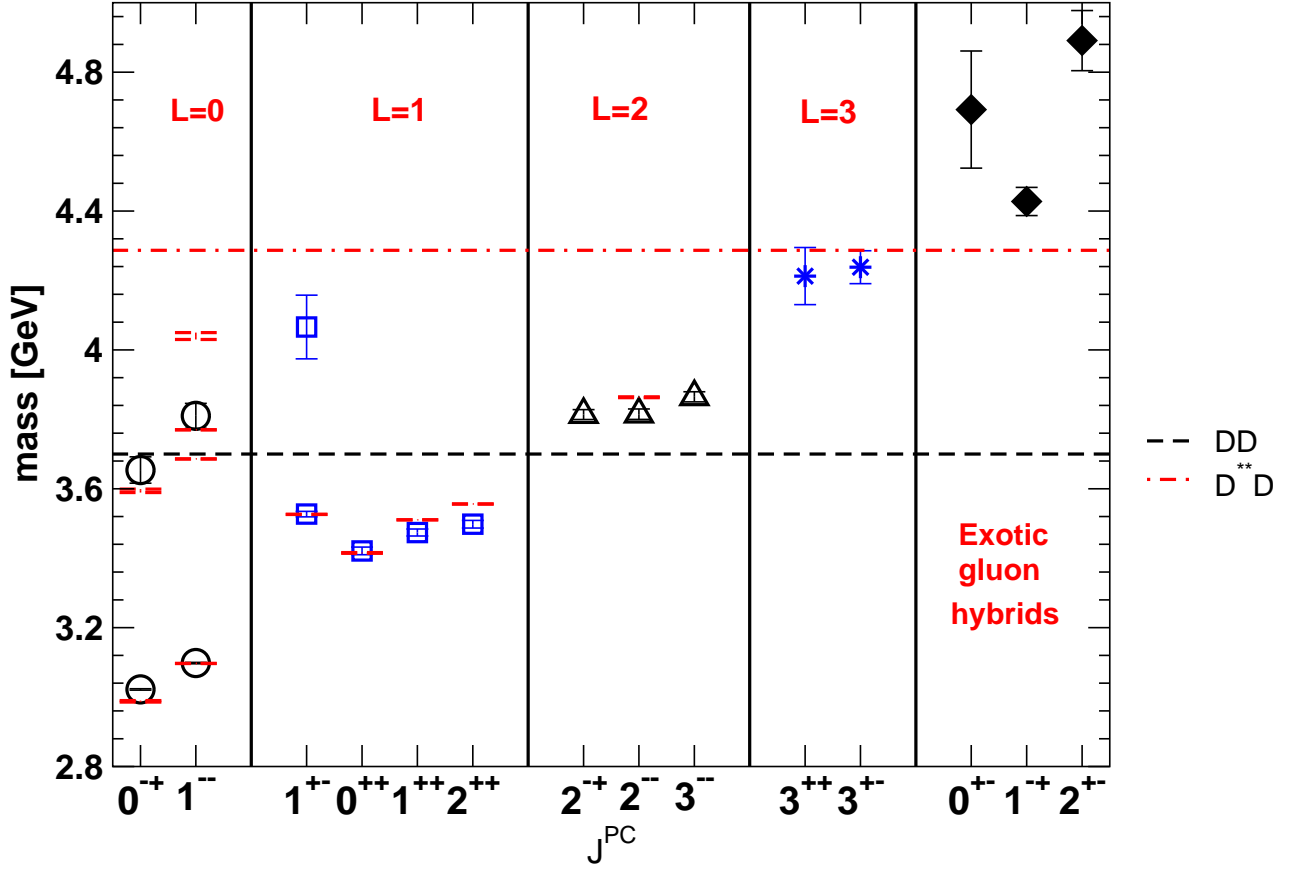}
\vskip0.5cm
\caption{Quenched charmonium spectrum. The experimental values are shown as 
	 short horizontal lines. The long horizontal dashed lines mark the $D\bar{D}$
	 and $D^{**}D$ thresholds.
	 The mesons (please refer to appendix \ref{appendix:mesonop} for our naming convention) plotted are: $\pi (0^{-+})$, $\rho (1^{--})$, $b_1 (1^{+-})$,
	$\rho \times \nabla\_A_1 (0^{++})$, $\rho \times \nabla\_T_1 (1^{++})$,  $\rho \times \nabla\_T_2 (2^{++})$,
	$\pi  \times D\_T_2 (2^{-+})$, $\rho \times D\_T_2 (2^{--})$, $\rho \times D\_A_2 (3^{--})$, 
        $a_1 \times D\_A_2 (3^{++})$,  $b_1 \times D\_A_2 (3^{+-})$,
	$a_1\times B\_A_1 (0^{+-})$, $\rho \times B\_T_1 (1^{-+})$, and $a_1 \times B\_T_2 (2^{+-})$.
	Some of the low-lying meson ($\pi (0^{-+})$, $\rho (1^{--})$, $b_1 (1^{+-})$) are taken 
	from Coumbia group's previous work \protect \cite{aniso:pchen}.
	The lattice scale is set by the $^1P_1-1S$ splitting. The numerical values are listed in table
        \ref{tab:result}.
        }
\label{fig:charm_spect}
\end{center}
\end{figure}

\begin{figure}[htb]
\begin{center}
  \epsfig{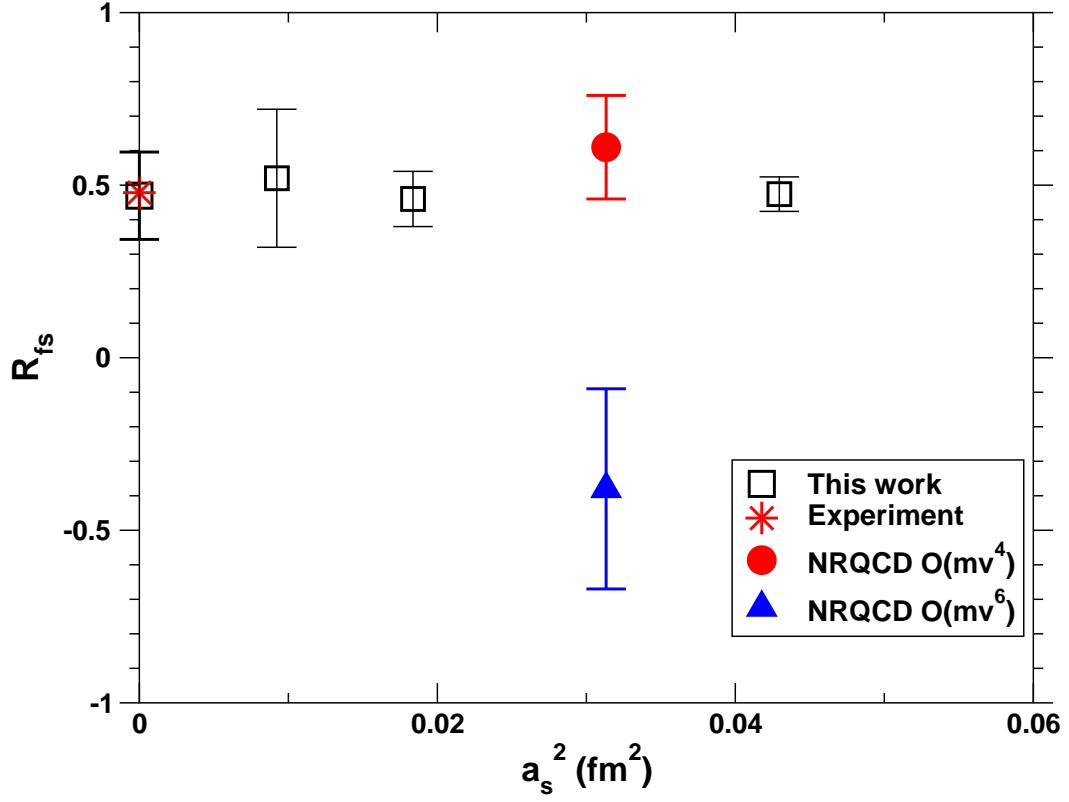}
\vskip0.5cm
\caption{ P-wave triplet splitting ratio $R_{fs}={(^3P_2-^3P_1) / (^3P_1-^3P_0)}$. Comparison
  is shown 
  with NRQCD results with relativistic corrections up to $O(mv^4)$ and $O(mv^6)$
  \protect \cite{thesis:manke}.
  }
\label{fig:3pj_sumamry}
\end{center}
\end{figure}

\begin{figure}[htb]
\begin{center}
\epsfig{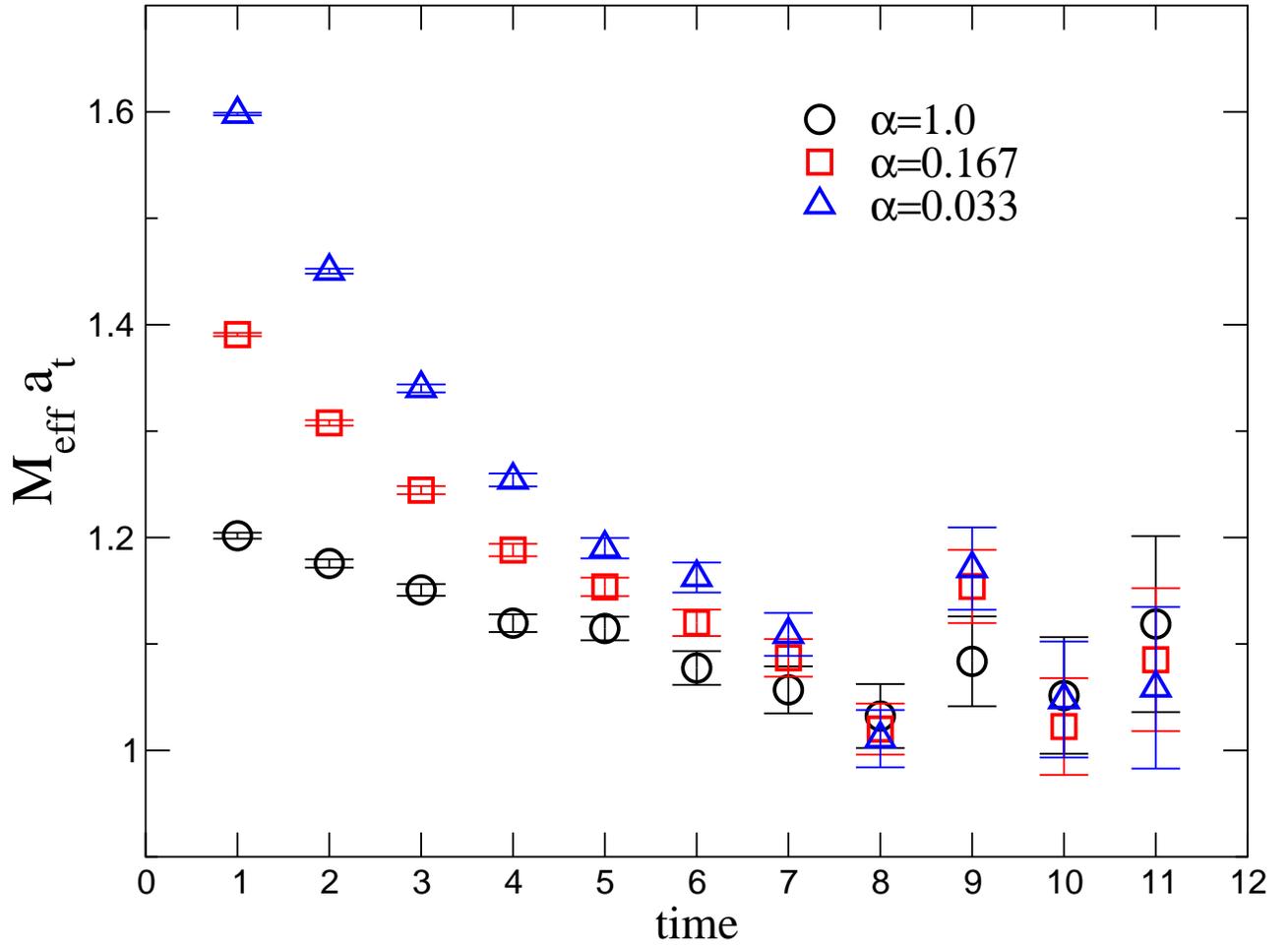}
\vskip0.5cm
\caption{ Effective mass of the $1^{-+}$ exotic meson  for $\beta=6.1$ and 
	$\xi=2$ with different APE-style fuzzing coeffients $\alpha$ as defined in equation \ref{eq:ape}.}
\label{fig:eff_mass_rhoxB_T1_b61}
\end{center}
\end{figure}

\begin{figure}[htb]
\begin{center}
\epsfig{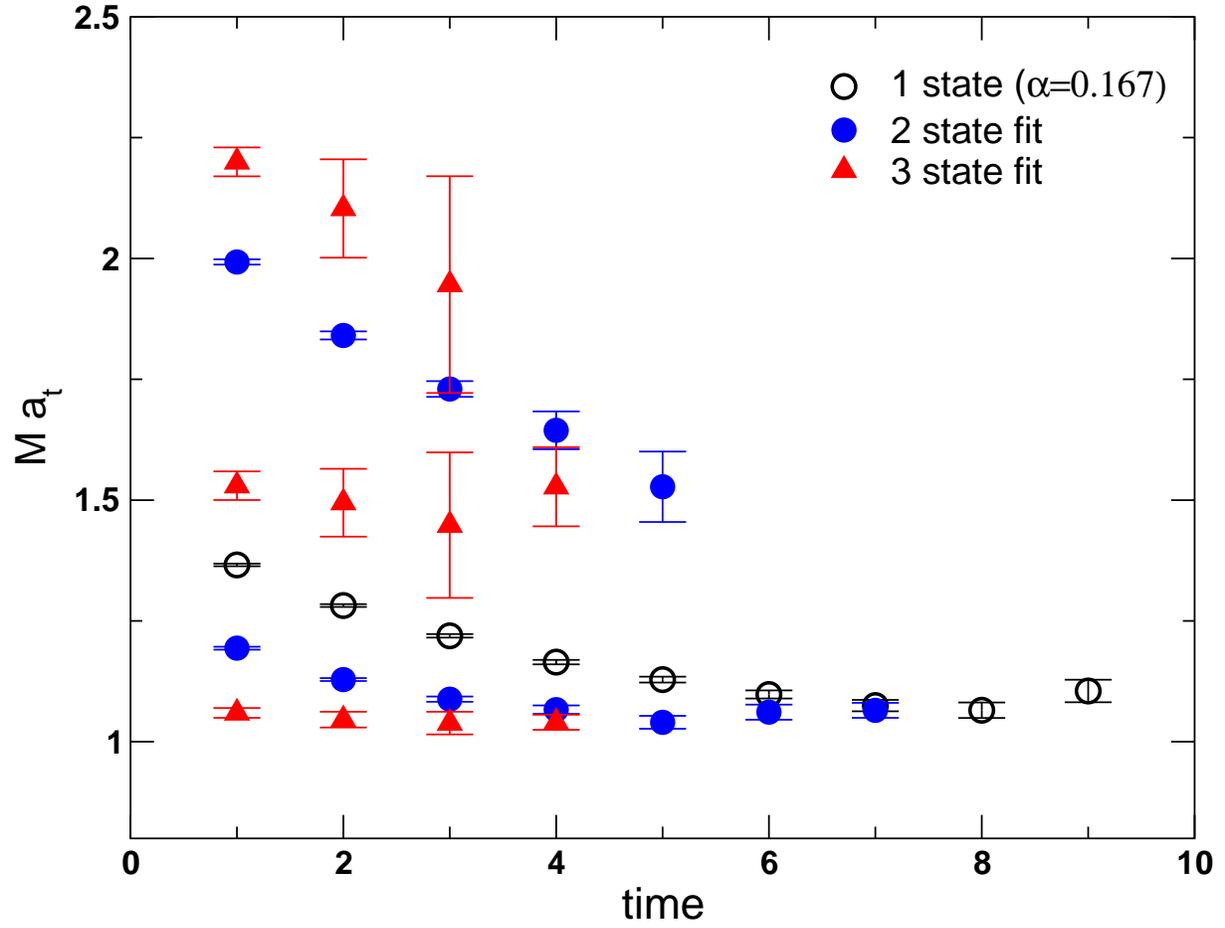}
\vskip0.5cm
\caption{ Masses from multi-exponential fits of $\rho \times B\_T_1$ propagator (the ground
          state is the exotic $1^{-+}$) for $\beta=6.1$ and 
	$\xi=2$.The upper band of solid circles and two added upper bands of solid triangles show
        the effective excited state masses.}
\label{fig:fit_rhoxB_T1_b61}
\end{center}
\end{figure}
\begin{figure}[htb]
\begin{center}
\epsfig{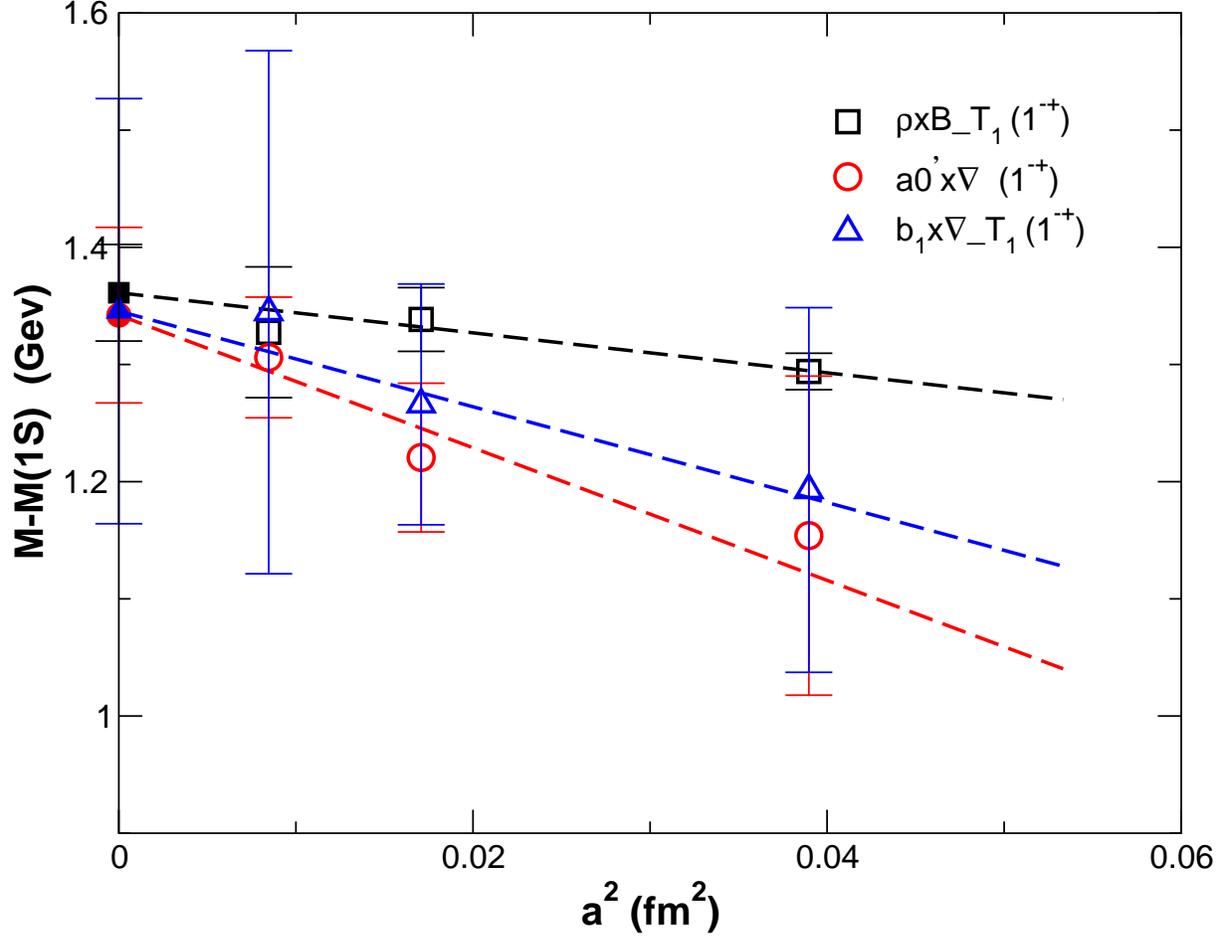}
\vskip0.5cm
\caption{ Continuum extrapolation (as outlined in section \ref{sec:cont_extrap}) of the exotic hybrid $1^{-+}$ (mass above $1S$) with 
	  different meson operators
	 ( $\rho \times B\_T_1$, $a_0^{\prime}\times \nabla$, and $b_1 \times \nabla\_T_1$).
	}
\label{fig:fit_compare_op_extrap_1-+}
\end{center}
\end{figure}

\begin{figure}[htb]
\begin{center}
\epsfig{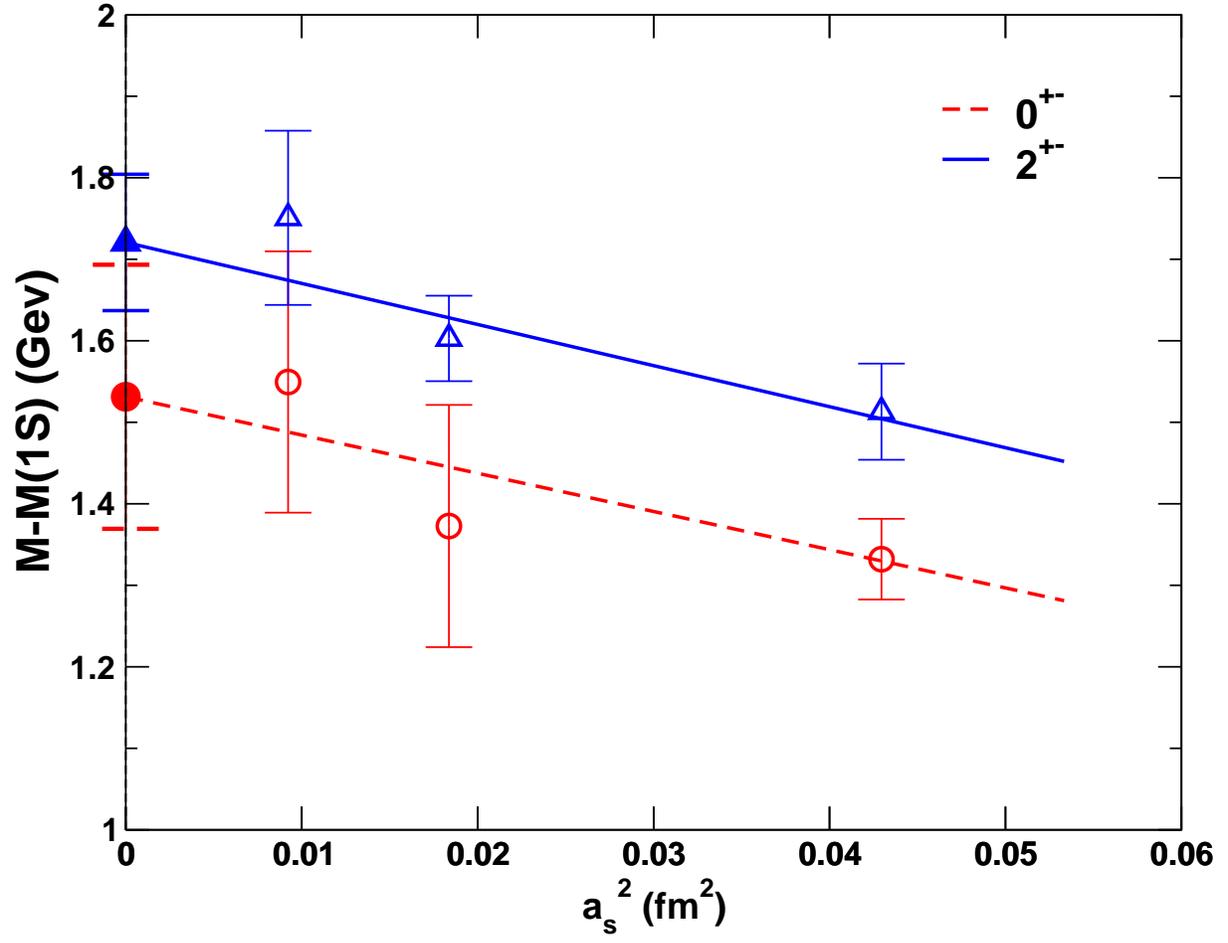}
\vskip0.5cm
\caption{ Continuum extrapolation of the exotic hybrid $0^{+-}$ ($a_1 \times B\_A_1$) and $2^{+-}$ ($a_1 \times B\_T_2$).
	}
\label{fig:fit_compare_op_extrap_0+-2+-}
\end{center}
\end{figure}

\begin{figure}[htb]
\begin{center}
\epsfig{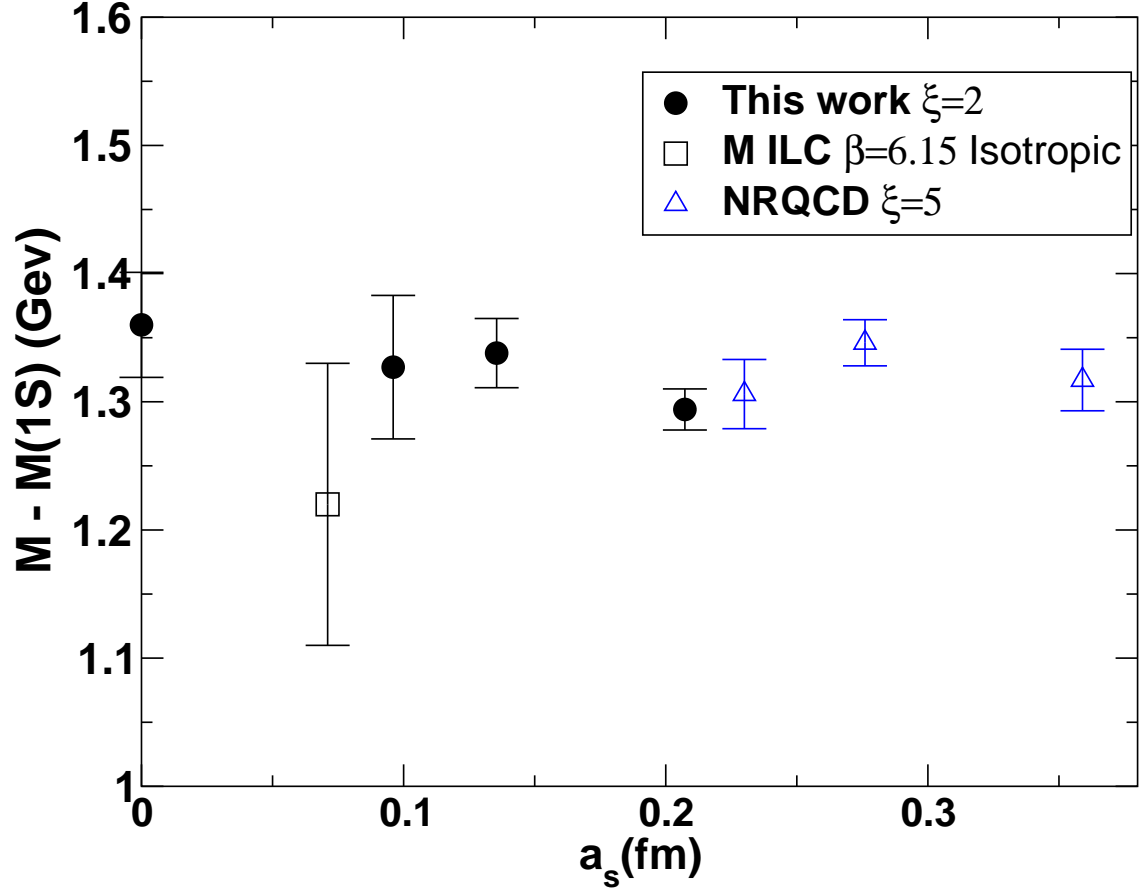}
\vskip0.5cm
\caption{ Comparison of charmonium hybrid meson $1^{-+}$ mass (relative
	  to the spin-averaged S-wave meson mass $1S$) as determined
	by this work, NRQCD (CP-PACS \protect \cite{CAPACS-manke-hybrid1})
	, and isotropic relativistic lattice QCD (MILC collaboration
	\protect \cite{milc_hybrid}). 
}
\label{fig:fit_compare_1-+_others}
\end{center}
\end{figure}

%% file: rel_charm.bbl
\begin{thebibliography}{10}

\bibitem{hybrid-review:page}
P.~R. Page,
\newblock {Nucl. Phys.} A {\bf 663}, 585 (2000).

\bibitem{cleo-C:stock2002}
H.~Stock,
\newblock (2002), hep-ex/0204015,
\newblock presented at the XXXVIIth Recontres de Moriond - QCD and High Energy
  Hadronic Interaction.

\bibitem{BES-charm:2001}
Z.~Zhao,
\newblock eConf {\bf C010430}, M10 (2001), hep-ex/0106028.

\bibitem{nrqcd:improve}
G.~P. Lepage {\em et~al.},
\newblock {Phys. Rev.} D {\bf 46}, 4052 (1992).

\bibitem{heavy:Davies}
C.~Davies,
\newblock {\em The Heavy Hadron Spectrum},
\newblock Lectures at Schladming Winter School 1997, hep-ph/9710394.

\bibitem{nrqcd:trottier-rel}
N.~H. Shakespeare and H.~D. Trottier,
\newblock {Phys. Rev.} D {\bf 58}, 034502 (1998).

\bibitem{nrqcd:trottier-spin}
H.~D. Trottier,
\newblock {Phys. Rev.} D {\bf 55}, 6844 (1997).

\bibitem{nrqcd:manke-bottom-1997}
UKQCD, T.~Manke, I.~T. Drummond, R.~R. Horgan, and H.~P. Shanahan,
\newblock Phys. Lett. {\bf B408}, 308 (1997).

\bibitem{nrqcd:Eicker-bottom-1998}
N.~Eicker {\em et~al.},
\newblock Phys. Rev. {\bf D57}, 4080 (1998).

\bibitem{Fermilab}
A.~X. El-Khadra, A.~S. Kronfeld, and P.~B. Mackenzie,
\newblock {Phys. Rev.} D {\bf 55}, 3933 (1997).

\bibitem{Fermilab:ukqcdmeson}
P.~Boyle,
\newblock {Nucl. Phys.} (Proc.Suppl.) {\bf 63}, 314 (1998).

\bibitem{aniso:karsch}
F.~Karsch,
\newblock {Nucl. Phys.} B {\bf 205}, 285 (1982).

\bibitem{aniso:glueball:1997}
C.~J. Morningstar and M.~J. Peardon,
\newblock Phys. Rev. {\bf D56}, 4043 (1997).

\bibitem{aniso:glueball:1999}
C.~J. Morningstar and M.~J. Peardon,
\newblock Phys. Rev. {\bf D60}, 034509 (1999).

\bibitem{aniso:thermo-Namekawa2001}
CP-PACS, Y.~Namekawa {\em et~al.},
\newblock Phys. Rev. {\bf D64}, 074507 (2001).

\bibitem{aniso:klassen-improv1997}
M.~G. Alford, T.~R. Klassen, and G.~P. Lepage,
\newblock Nucl. Phys. {\bf B496}, 377 (1997).

\bibitem{aniso:klassen-latproc}
T.~R. Klassen,
\newblock {Nucl. Phys.} (Proc.Suppl.) {\bf 73}, 918 (1999).

\bibitem{aniso:pchen}
P.~Chen,
\newblock {Phys. Rev.} D {\bf 64}, 034509 (2001).

\bibitem{aniso:cppacs-charm-lat00}
A.~{Ali Khan {\em et al.} CP-PACS Collaboration},
\newblock {Nucl. Phys.} (Proc.Suppl.) {\bf 94}, 325 (2001),
\newblock hep-lat/0011005.

\bibitem{own-bottom-paper}
X.~Liao and T.~Manke,
\newblock {Phys. Rev.} D {\bf 65}, 074508 (2002).

\bibitem{hybrid:bornapprox}
P.~Hasenfratz, R.~R. Horgan, J.~Kuti, and J.~M. Richard,
\newblock Phys. Lett. {\bf B95}, 299 (1980).

\bibitem{bagmodel:barnes}
T.~Barnes,
\newblock Nucl. Phys. {\bf B158}, 171 (1979).

\bibitem{bagmodel:charm}
P.~Hasenfratz, R.~R. Horgan, J.~Kuti, and J.~M. Richard,
\newblock Phys. Lett. {\bf B95}, 299 (1980).

\bibitem{flux_tube}
N.~Isgur and J.~Paton,
\newblock {Phys. Rev.} D {\bf 31}, 2910 (1985).

\bibitem{sumrule:1982}
I.~I. Balitsky, D.~Diakonov, and A.~V. Yung,
\newblock {Phys. Lett.} B {\bf 112}, 71 (1982).

\bibitem{sumrule:heavy}
J.~I. Latorre, S.~Narison, P.~Pascual, and R.~Tarrach,
\newblock Phys. Lett. {\bf B147}, 169 (1984).

\bibitem{NRQCDHybrids-ukqcd}
T.~Manke, I.~T. Drummond, R.~R. Horgan, and H.~P. Shanahan,
\newblock {Phys. Rev.} D {\bf 57}, 3829 (1998),
\newblock hep-lat/9710083.

\bibitem{milc_hybrid}
C.~W. Bernard {\em et~al.},
\newblock {Phys. Rev.} D {\bf 56}, 7039 (1997).

\bibitem{exotic-review-Kuti1998}
J.~Kuti,
\newblock Nucl. Phys. Proc. Suppl. {\bf 73}, 72 (1999).

\bibitem{own-lat00}
P.~Chen, X.~Liao, and T.~Manke,
\newblock {Nucl. Phys.} (Proc.Suppl.) {\bf 94}, 342 (2001).

\bibitem{aniso:aoki}
S.~Aoki, Y.~Kuramashi, and S.~Tominaga,
\newblock (2001),
\newblock hep-lat/0107009.

\bibitem{APEFuzzing}
M.~{Albanese {\em et al.} - APE Collaboration},
\newblock {Phys. Lett.} B {\bf 192}, ~163 (1987).

\bibitem{sommerscale}
R.~Sommer,
\newblock {Nucl. Phys.} B {\bf 411}, 839 (1994).

\bibitem{aniso:cppacs-charm}
M.~{Okamoto {\em et al.} CP-PACS Collaboration},
\newblock (2001),
\newblock hep-lat/0112020.

\bibitem{thesis:manke}
T.~Manke,
\newblock {\em The Spectroscopy of Heavy Quark Systems},
\newblock PhD thesis, Darwin College, Cambridge, 2000.

\bibitem{CAPACS-manke-hybrid1}
T.~Manke {\em et~al.},
\newblock {Phys. Rev. Lett.} {\bf 82}, 4396 (1999).

\bibitem{hybrid-potential-lat}
K.~Juge, J.~Kuti, and C.~Morningstar,
\newblock {Nucl. Phys.} (Proc.Suppl.) {\bf 63}, 326 (1998).

\bibitem{sum_rules_hybrid}
J.~G. {\it et al.},
\newblock Nucl. Phys. {\bf {\bf B}284}, ~674 (1987).

\bibitem{hybrid-decay:page}
P.~R. Page, E.~S. Swanson, and A.~P. Szczepaniak,
\newblock {Phys. Rev.} D {\bf 59}, 034016 (1999).

\bibitem{hybrid-decay-SS-page}
P.~R. Page,
\newblock {Phys. Lett.} B {\bf 402}, 183 (1997).

\bibitem{Drummond:1999db}
I.~T. Drummond {\em et~al.},
\newblock Phys. Lett. {\bf B478}, 151 (2000), hep-lat/9912041.

\bibitem{ludmila-aniso}
L.~Levkova and T.~Manke,
\newblock {Nucl. Phys.} (Proc.Suppl.) {\bf 106}, 218 (2002).

\end{thebibliography}
